\documentclass[a4paper,11pt]{article}
\usepackage{jheppub}
\usepackage{CJK}
\usepackage{tikz}
\usepackage{float,graphbox}
\usepackage[hang]{footmisc}
\makeatletter
\ifFN@para
\else
  \long\def\@makefntext#1{%
    \ifFN@hangfoot
      \bgroup
      \setbox\@tempboxa\hbox{%
        \ifdim\footnotemargin>0pt
          \hb@xt@\footnotemargin{\@makefnmark\hss}%
        \else
          \@makefnmark\hskip-\footnotemargin      
        \fi
      }%
      \leftmargin\wd\@tempboxa
      \rightmargin\z@
      \linewidth \columnwidth
      \advance \linewidth -\leftmargin
      \parshape \@ne \leftmargin \linewidth
      \footnotesize
      \@setpar{{\@@par}}%
      \leavevmode
      \llap{\box\@tempboxa}%
      \parskip\hangfootparskip\relax
      \parindent\hangfootparindent\relax
    \else
      \parindent1em
      \noindent
      \ifdim\footnotemargin>\z@
        \hb@xt@ \footnotemargin{\hss\@makefnmark}%
      \else
        \ifdim\footnotemargin=\z@
          \llap{\@makefnmark}%
        \else
          \llap{\hb@xt@ -\footnotemargin{\@makefnmark\hss}}%
        \fi
      \fi
    \fi
    \footnotelayout#1%
    \ifFN@hangfoot
      \par\egroup
    \fi
  }
\fi
\makeatother

\begin{document}

\begin{CJK*}{UTF8}{}
\CJKfamily{gbsn}

\title{The Cusp Limit of Correlators and a New Graphical Bootstrap for Correlators/Amplitudes to Eleven Loops}

\author[a,b,c]{Song He (何颂)}
\emailAdd{songhe@itp.ac.cn}
\author[a]{Canxin Shi (施灿欣)}
\emailAdd{shicanxin@itp.ac.cn}
\author[a,d]{Yichao Tang (唐一朝)}
\emailAdd{tangyichao@itp.ac.cn}
\author[a,d]{Yao{-}Qi Zhang (张耀奇)}
\emailAdd{zhangyaoqi@itp.ac.cn}

\affiliation[a]{CAS Key Laboratory of Theoretical Physics, Institute of Theoretical Physics, Chinese Academy of Sciences, Beijing 100190, China}
\affiliation[b]{School of Fundamental Physics and Mathematical Sciences, Hangzhou Institute for Advanced Study and ICTP-AP, UCAS, Hangzhou 310024, China}
\affiliation[c]{Peng Huanwu Center for Fundamental Theory, Hefei 230026, China}
\affiliation[d]{School of Physical Sciences, University of Chinese Academy of Sciences, No.19A Yuquan Road, Beijing 100049, China}

\abstract{
We consider the universal behavior of half-BPS correlators in $\mathcal N=4$ super-Yang-Mills in the cusp limit where two consecutive separations $x_{12}^2,x_{23}^2$ become lightlike. Through the Lagrangian insertion procedure, the Sudakov double-logarithmic divergence of the $n$-point correlator is related to the $(n+1)$-point correlator where the inserted Lagrangian ``pinches'' to the soft-collinear region of the cusp.
We formulate this constraint as a new {\it graphical rule} for the $f$-graphs of the four-point correlator, which turns out to be the most constraining rule known so far. By exploiting this single graphical rule, we bootstrap the planar integrand of the four-point correlator up to ten loops ($n=14$) and fix all 22024902 but one coefficient at eleven loops ($n=15$); the remaining coefficient is then fixed using the triangle rule. We verify the ``Catalan conjecture" for the coefficients of the family of $f$-graphs known as ``anti-prisms" where the coefficient of the twelve-loop ($n=16$) anti-prism is found to be $-42$ by a local analysis of the bootstrap equations. 
We also comment on the implication of our graphical rule for the non-planar contributions.
}

\maketitle
\end{CJK*}

\section{Introduction}
Recent years have witnessed enormous progress in the study of scattering amplitudes, not only for developing powerful new computational tools, but also for revealing hidden simplicity and mathematical structures of quantum field theory. Needless to say, the most symmetric of all four-dimensional gauge theories, the ${\cal N}=4$ supersymmetric Yang-Mills (SYM) theory, has played a central role in this story. In particular, correlation functions of half-BPS operators (the simplest local operators in the theory) have been of enormous interest because they stand at the crossroads of perturbative CFT, integrability, gauge/string duality and conformal bootstrap.

Remarkably, there are two seemingly unrelated but equally fascinating relations between half-BPS correlators and scattering amplitudes in two different theories (see the nice review~\cite{Heslop:2022xgp} and references therein). At strong coupling via AdS/CFT, the half-BPS correlators are dual to type-IIB supergravity amplitudes in string theory on AdS$_5\times S^5$ background, which provides invaluable theoretical data for the study of quantum gravity. At weak coupling, we have the famous correlator/amplitude duality, which is really a triality relating a certain lightlike limit of the correlators to null polygonal Wilson loops~\cite{Alday:2010zy,Adamo:2011dq}, which in the planar limit equals the square of scattering amplitudes in ${\cal N}=4$ SYM~\cite{Alday:2007hr,Drummond:2007aua,Brandhuber:2007yx,Bern:2008ap,Drummond:2008aq,Eden:2010zz,Arkani-Hamed:2010zjl,Mason:2010yk,Eden:2010ce,Caron-Huot:2010ryg,Eden:2011yp,Eden:2011ku}. This duality holds at the integrand level, and has provided important insights and rich data for both correlators and amplitudes. Take for example the simplest four-point correlator of the stress-tensor supermultiplet, in various lightlike limits, the planar $\ell$-loop integrand already contains the $\ell$-loop integrand of four-point amplitudes, and $(\ell-1)$-loop five-point amplitudes (squared) {\it etc.}~\cite{Eden:2010zz, Ambrosio:2013pba, Heslop:2018zut}, and in fact the most accurate perturbative data for integrands of amplitudes are obtained in this way from that of the four-point correlators! The knowledge that planar correlators are related to squared amplitudes also imposes powerful constraints on the correlators, which allows for impressive perturbative results. 

Let us take a quick look at the history of perturbative study of four-point correlators of the stress-tensor supermultiplet; for results of higher Kaluza-Klein modes and higher points, see~\cite{Chicherin:2015edu,Chicherin:2018avq,Caron-Huot:2021usw,Bargheer:2022sfd,Caron-Huot:2023wdh}. Already in the early days of AdS/CFT, one- and two-loop correlators (both integrands and integrated results) have been determined~\cite{Gonzalez-Rey:1998wyj, Eden:1998hh, Eden:1999kh, Eden:2000mv, Bianchi:2000hn}, but it has to wait until 2011 before the planar integrand of the correlator was determined for three loops~\cite{Eden:2011we}, which was followed quickly by results for four, five, six~\cite{Eden:2012tu, Ambrosio:2013pba} and even seven loops~\cite{Ambrosio:2013pba} (the integrated results are known up to three loops~\cite{Drummond:2013nda}). The key breakthrough was the discovery of a hidden permutation symmetry for the integrand of the correlator~\cite{Eden:2011we}, which when combined with correlator/amplitude duality and the soft-collinear bootstrap~\cite{Bourjaily:2011hi}, has enabled the determination of eight-loop integrands (for both correlators and amplitudes)~\cite{Bourjaily:2015bpz}. 

As we will review shortly, the $\ell$-loop integrand of four-point correlators can be computed as the Born-level correlator with $4$ operators and $\ell$ chiral Lagrangian insertions~\cite{Eden:2011yp}, which can be normalized to enjoy the symmetry under full $S_{4{+}\ell}$ permutation group; a basis of such functions can be nicely represented as {\it $f$-graphs} with $4{+} \ell$ vertices (each with net valency $4$ due to conformal symmetry), which can be easily enumerated ({\it e.g.} to at least twelve loops in the planar limit, see Table~\ref{tab:stats}). The remaining question becomes how to determine the coefficients of $f$-graphs, for which the soft-collinear bootstrap has succeeded up to eight loops~\cite{Bourjaily:2015bpz}. To proceed to even higher loops, various {\it graphical rules} were proposed in~\cite{Bourjaily:2016evz}, which have greatly improved the efficiency of such a bootstrap (as opposed to rules/constraints on rational functions); based on universal behavior of the correlator known as the {\it triangle rule}, as well as consistency of the correlator/amplitude duality known as the {\it square rule} and {\it pentagon rule}, the authors have determined the integrands up to ten loops! Much less is known beyond the planar limit: the first non-trivial non-planar correction comes in four loops, which is the only data available so far and follows from a heroic calculation~\cite{Eden:2012tu, Fleury:2019ydf} based on Feynman diagrams (in twistor space)~\cite{Chicherin:2014uca}. 

In this paper, we push the frontier even further, up to eleven loops in the planar limit, by considering the universal behavior of the correlator in the so-called ``cusp limit" and deriving a new graphical rule on $f$-graphs from this constraint. 

The cusp limit of the half-BPS correlators (in Minkowskian signature) arises from the limit where two consecutive separations $x_{12}^2,x_{23}^2$ become lightlike, forming a Wilson line with a single cusp. The cusped Wilson line has the familiar Sudakov double-logarithmic divergence. According to the Lagrangian insertion procedure (which we review in appendix~\ref{app:lagrangian}, collecting the main results in the literature and addressing many subtleties in great detail), this divergence arises from the soft-collinear region where the inserted Lagrangian ``pinches" to the cusp. This imposes a constraint relating the $(n+1)$-point Born-level correlator (or the $(n-4)$-loop integrand of the four-point correlator) to the $n$-point Born-level correlator, which we derive in section~\ref{sec:cusp}. Since our argument is essentially one-loop, we are able to avoid analyzing the complicated sub-divergences in multi-loop integrals, putting the argument on solid grounds.

We then reformulate the cusp limit constraint as a new graphical rule in section~\ref{sec:rule}, which pinches the waist of double-triangle subgraphs in $(n+1)$-point $f$-graphs to arrive at $n$-point $f$-graphs. This way, we obtain a set of equations relating the coefficients $\{c_i^{(n+1)}\}$ and $\{c_j^{(n)}\}$, which we call the $(n+1)\to n$ bootstrap equations. Note that this is in spirit similar to the triangle rule which arises from the OPE limit in Euclidean signature, which we rederive in appendix~\ref{app:triangle} also using an essentially one-loop argument and avoiding complications related to sub-divergences in multi-loop integrals. 

It turns out that our new graphical rule is the most constraining graphical rule known so far, which we demonstrate in section~\ref{sec:boots}. It allows us to fix all coefficients of planar $f$-graphs up to ten loops and all but one out of 22024902 coefficients at eleven loops ($n=15$ vertices), without using any other graphical rules! We fix the remaining eleven-loop coefficient with the help of the triangle rule, arriving at a unique solution which can be found in the ancillary file\footnote{\texttt{ancillary\_file.zip} can be downloaded \href{https://drive.google.com/file/d/1SfFY-eDNMIM6zmhmOwXmAY5cKhdsc_fr/view?usp=sharing}{here},
which contains the planar $f$-graphs up to 15 points and their coefficients, all $f$-graphs up to 9 points, and a Mathematica notebook \texttt{demo.nb} containing some demonstrations and the bootstrap equations involving the 16-point anti-prism. Note that although the \texttt{.zip} file is only 100MB, the uncompressed data file takes up around 1GB. We thank Jacob Bourjaily for inventing a method to compress such a large amount of $f$-graph data.}. In fact, this last coefficient corresponds to the unique 15-point planar $f$-graph (the 15-point chopped-prism) that has no double-triangle subgraphs, which makes it invisible in the $15\to14$ bootstrap equations.

Actually, invisible $f$-graphs already have appeared at $n=12$ (the 12-point chopped-prism), although there are no invisible $f$-graphs at $n=13,14$. How is it possible that our new graphical rule fixes the 12-point ansatz?

The 12-point chopped-prism is indeed invisible in the $12\to11$ bootstrap equations. However, it is visible in and fixed by the $13\to12$ bootstrap equations. In fact, we observe that \textbf{the $(n+1)\to n$ bootstrap equations fixes both $\{c_j^{(n)}\}$ and the visible $\{c_i^{(n+1)}\}$}, at least up to $n=14$. We stress that this is not the familiar recursive idea, where one attempt to fix $\{c_j^{(n)}\}$ given $\{c_k^{(n-1)}\}$. We can directly start with the $(n+1)\to n$ bootstrap equations and obtain $\{c_j^{(n)}\}$, even without knowing $\{c_k^{(n-1)}\}$. For this reason, we strongly suspect that the last 15-point coefficient can also be fixed by the $16\to15$ bootstrap equations. However, since there are 619981403 planar $f$-graphs at $n=16$, and finding the complete $16\to15$ bootstrap equations would take too long.

Note that similar to the triangle rule, our new graphical rule is oblivious to plane embeddings (unlike the square and pentagon rules) and more importantly, it also applies to non-planar contributions. In section~\ref{sec:nonplanar}, we will present some preliminary results of applying the new rule to correlator integrands beyond the planar limit at three to five loops ($n=7,8,9$).

A nice property of the bootstrap equations is that they are surprisingly easy to solve: most equations are very short, of the form $c_i^{(n)} =c_i^{(n{-}1)}$ or $c_i^{(n)}=0$. This is because our graphical rule is many-to-one; each bootstrap equation relates many $c_i^{(n)}$ to a single $c_j^{(n-1)}$ (or 0). It is then natural to study ``local'' subsystems of the bootstrap equations, which involves only a small number of $f$-graphs. If such a closed subsystem exists, it is usually much cheaper to study these bootstrap equations locally than to obtain the complete set of bootstrap equations. In section~\ref{sec:catalan}, we demonstrate the power of such local analysis by considering the special anti-prism $f$-graphs. Specifically, for each even $n$, there is a unique anti-prism graph which seems to always have the largest coefficient (in magnitude): for $n=8,10,12,14$, the coefficients are $-1, 2, -5, 14$ respectively. Moreover, these values coincide with the first few Catalan numbers, and a natural conjecture proposed in~\cite{Bourjaily:2016evz} is that the next instance at twelve loops ($n=16$) will have coefficient $-42$. For general even $n$, we find a beautiful linear equation when pinching the anti-prism, which is the only $n\to(n-1)$ bootstrap equation involving the anti-prism coefficient. It turns out to be very easy to find a closed subsystem of $n\to(n-1)$ bootstrap equations that fully determine the $f$-graph coefficients that appear in that equation, which enables us to solve for the anti-prism coefficient locally. In the first version of the manuscript, we have mistakenly solved the system, and a corrected result indicates that this ``Catalan conjecture" still holds at twelve loops. 
Additionally, the aforementioned invisible 12-point chopped-prism also lives in a simple local subsystem, which we describe in section~\ref{sec:discussions}; there we discuss the possible existence of a similar local subsystem involving the invisible 15-point chopped-prism, as well as various other interesting future directions.

\section{The Cusp Limit of Correlators}\label{sec:cusp}

Consider the half-BPS operators in the stress tensor supermultiplet in $\mathcal N=4$ SYM with gauge group $SU(N_c)$:
\begin{equation}
    O(x,t):=t^It^J\mathop{\rm tr}(\phi^I\phi^J)(x).
\end{equation}
Here, $I,J=1,\cdots,6$ are vector indices of the $\mathfrak{so}(6)$ R-symmetry, and the auxiliary vector $t^I$ satisfies $t^2=0$, projecting out the $\mathbf{20}'$ representation. We are interested in the connected correlator of four such operators:
\begin{equation}
\begin{split}
    \langle O(x_1,t_1)O(x_2,t_2)O(x_3,t_3)O(x_4,t_4)\rangle_{\rm conn}&=\mathcal R(1,2,3,4)\times G(u,v),\\ u=\frac{x_{12}^2x_{34}^2}{x_{13}^2x_{24}^2},\quad v=&\frac{x_{14}^2x_{23}^2}{x_{13}^2x_{24}^2}. 
\end{split}
\end{equation}
In the above expression, we have factored out all R-symmetry dependence and conformal weights into $\mathcal R(1,2,3,4)$, which is possible because $\mathcal R$ is the unique superconformal invariant at four points (see the nice review~\cite{Heslop:2022xgp} for details). Then, without loss of generality, we may consider the component $\langle\mathcal O(x_1)\bar{\mathcal O}(x_2)\mathcal O(x_3)\bar{\mathcal O}(x_4)\rangle$, where
\begin{equation}
    \begin{aligned}
        \mathcal O:=\mathop{\rm tr}(\varphi^2),&\quad\varphi:=\frac{\phi^1+i\phi^2}{\sqrt2},\\
        \bar{\mathcal O}:=\mathop{\rm tr}(\bar\varphi^2),&\quad\bar\varphi:=\frac{\phi^1-i\phi^2}{\sqrt2}.
    \end{aligned}
\end{equation}
For such a choice, $t_{12}=t_{23}=t_{34}=t_{14}=1$ and $t_{13}=t_{24}=0$, and
\begin{equation}
    \langle\mathcal O(x_1)\bar{\mathcal O}(x_2)\mathcal O(x_3)\bar{\mathcal O}(x_4)\rangle_{\rm conn}=\frac{N_c^2-1}{(4\pi^2)^4}\frac1{x_{12}^2x_{23}^2x_{34}^2x_{14}^2}\times G(u,v).
\end{equation}
For $g_{\rm YM}\ll1$, $G(u,v)$ admits the following perturbative expansion:
\begin{equation}
    G(u,v)=1+\sum_{\ell=1}^\infty a^\ell G^{(\ell)}(u,v),\quad a:=\frac{g_{\rm YM}^2N_c}{4\pi^2}.
\end{equation}

The more general class of correlators with (chiral, on-shell) Lagrangian insertions will be of great use to us:
\begin{equation}
    \langle\mathcal O_1\bar{\mathcal O}_2\mathcal O_3\bar{\mathcal O}_4(g_{\rm YM}^{-2}\mathcal L_5)\cdots(g_{\rm YM}^{-2}\mathcal L_n)\rangle_{\rm conn}=\sum_{\ell=0}^\infty(g_{\rm YM}^2)^\ell\langle\mathcal O_1\bar{\mathcal O}_2\mathcal O_3\bar{\mathcal O}_4(g_{\rm YM}^{-2}\mathcal L_5)\cdots(g_{\rm YM}^{-2}\mathcal L_n)\rangle_{\rm conn}^{(\ell)},
\end{equation}
where the subscript $_i$ denotes an operator at $x_i$. The chiral, on-shell Lagrangian is defined as~\cite{Eden:2011yp}
\begin{equation}
    \mathcal L=\mathop{\rm tr}\left(-\frac12F_{\alpha\beta}F^{\alpha\beta}+\sqrt2g_{\rm YM}\psi^{A\alpha}[\phi_{AB},\psi^B_\alpha]-\frac{g_{\rm YM}^2}8[\phi^{AB},\phi^{CD}][\phi_{AB},\phi_{CD}]\right).
\end{equation}
A crucial property of $\mathcal N=4$ SYM is that the perturbation series can be computed using Lagrangian insertions\footnote{See appendix~\ref{app:lagrangian} for details.}:
\begin{equation}\label{eq:insertionM}
    \begin{aligned}
        \langle\mathcal O_1\bar{\mathcal O}_2\mathcal O_3\bar{\mathcal O}_4&(g_{\rm YM}^{-2}\mathcal L_5)\cdots(g_{\rm YM}^{-2}\mathcal L_n)\rangle_{\rm conn}^{(\ell)}\\
        &=\frac{(-i)^\ell}{\ell!}\int{\rm d}^4x_{n+1}\cdots{\rm d}^4x_{n+\ell}\langle\mathcal O_1\bar{\mathcal O}_2\mathcal O_3\bar{\mathcal O}_4(g_{\rm YM}^{-2}\mathcal L_5)\cdots(g_{\rm YM}^{-2}\mathcal L_{n+\ell})\rangle_{\rm conn}^{(0)}.
    \end{aligned}
\end{equation}
The Born-level correlator takes the following form:
\begin{equation}
    \langle\mathcal O_1\bar{\mathcal O}_2\mathcal O_3\bar{\mathcal O}_4(a^{-1}\mathcal L_5)\cdots(a^{-1}\mathcal L_{4+\ell})\rangle_{\rm conn}^{(0)}=\frac{N_c^2-1}{(4\pi^2)^{4+\ell}}\frac1{x_{12}^2x_{23}^2x_{34}^2x_{14}^2}2\xi_4F_{4+\ell}(x_1,\cdots,x_{4+\ell}),
\end{equation}
where $\xi_4:=x_{13}^4x_{24}^4x_{12}^2x_{23}^2x_{34}^2x_{14}^2$ and $F_{4+\ell}$ is a rational function that (a) has at most simple poles\footnote{This is not merely an assumption, but is required by the form of OPEs. For details, see the discussion around eq.(3.10) in~\cite{Green:2020eyj}.}, (b) is invariant under arbitrary permutation of $x_1,\cdots,x_{4+\ell}$, and (c) has conformal weight 4 with respect to each $x_i$~\cite{Eden:2011we}. Therefore, after Wick rotating to Euclidean signature using ${\rm d}^4x^{(M)}=-i{\rm d}^4x^{(E)}$, we see that $F_{4+\ell}$ provides the integrand of $G^{(\ell)}(u,v)$:
\begin{equation}
    G^{(\ell)}(u,v)=\frac{2\xi_4}{\ell!}\int\frac{{\rm d}^4 x_5 \cdots {\rm d}^4 x_{4+\ell}}{(-4 \pi^2)^{\ell}}
    F_{4+\ell}.
\end{equation}

In~\cite{Alday:2010zy}, the authors presented a general argument for the correlator / Wilson loop duality, which states that the leading-power contribution of $\langle\mathcal O_1\bar{\mathcal O}_2\cdots\mathcal O_{2n-1}\bar{\mathcal O}_{2n}\rangle$ in the null polygon limit $x_{i,i+1}^2\to0$ is a null polygonal Wilson loop. The key fact needed in their argument is that in the presence of background gauge fields, the scalar propagator reads
\begin{equation}
    \langle\varphi(x)\bar\varphi(y)\rangle_A=\frac1{-4\pi^2(x-y)^2}{\rm P}\exp ig_{\rm YM}\int_y^xA_\mu(z){\rm d}z^\mu+\text{regular}=:\frac{[x,y]}{-4\pi^2(x-y)^2}+\text{regular},
\end{equation}
where we have denoted a straight Wilson line by $[x,y]$. Let us consider a more general ``cusp limit'' where only $x_{12}^2,x_{23}^2\to0$, forming a single cusp. Applying the same argument,
\begin{equation}\label{eq:master}
    \begin{aligned}
        &\langle\mathcal O_1\bar{\mathcal O}_2\mathcal O_3\bar{\mathcal O}_4(a^{-1}\mathcal L_5)\cdots(a^{-1}\mathcal L_n)\rangle_{\rm conn}\\
        =\,&\frac1{(4\pi^2)^2x_{12}^2x_{23}^2}\langle\mathop{\rm tr}\left(\varphi_1[x_1,x_2][x_2,x_3]\varphi_3\right)\bar{\mathcal O}_4(a^{-1}\mathcal L_5)\cdots(a^{-1}\mathcal L_n)\rangle_{\rm conn}+\text{regular}.
    \end{aligned}
\end{equation}
Consider the ratio of the order-$a^1$ and the order-$a^0$ contributions of the above equation. On the right-hand side, renormalizing the cusped Wilson line leads to the familiar Sudakov double-log factor:
\begin{equation}
    \frac{\text{RHS of eq.\eqref{eq:master}}|_{a^1}}{\text{RHS of eq.\eqref{eq:master}}|_{a^0}}=-\frac{\Gamma_{\rm cusp}^{{\rm adj.}}|_{a^1}}4\log x_{12}^2\log x_{23}^2+\text{other}.
\end{equation}
In particular, the ``other'' terms do not contribute to the $\log x_{12}^2\log x_{23}^2$ divergence. On the left-hand side, since loop corrections are computed with Lagrangian insertion,
\begin{equation}
    \frac{\text{LHS of eq.\eqref{eq:master}}|_{a^1}}{\text{LHS of eq.\eqref{eq:master}}|_{a^0}}=\frac{-i}{4\pi^2}\frac{\int{\rm d}^4x_{n+1}^{(M)}\,F_{n+1}}{F_n}.
\end{equation}
Putting the pieces together and moving around some terms, we get
\begin{equation}\label{eq:div}
    \left.\frac{-i}{4\pi^2}\lim_{x_{12}^2,x_{23}^2\to0}\int{\rm d}^4x_{n+1}^{(M)}\,x_{12}^2x_{23}^2F_{n+1}\right|_{\substack{\text{double-log}\\\text{divergence}}}=-\frac{\Gamma_{\rm cusp}^{{\rm adj.}}|_{a^1}}4\log x_{12}^2\log x_{23}^2\times\lim_{x_{12}^2,x_{23}^2\to0}x_{12}^2x_{23}^2F_n.
\end{equation}
On the left-hand side, we have a one-loop Feynman integral and the double-log divergence can only arise from the soft-collinear region $x_{n+1}\approx x_2$. We can extract the divergence as follows:
\begin{align*}
    &\text{LHS of eq.\eqref{eq:div}}\\
    =\,&\frac{-i}{4\pi^2}\left.\lim_{x_{12}^2,x_{23}^2\to0}\int\frac{x_{13}^2\,{\rm d}^4x_{n+1}^{(M)}}{x_{1,n+1}^2x_{2,n+1}^2x_{3,n+1}^2}\right|_{\substack{\text{double-log}\\\text{divergence}}}\times\lim_{\substack{x_{12}^2,x_{23}^2\to0\\x_{n+1}\to x_2}}\frac{x_{12}^2x_{23}^2x_{1,n+1}^2x_{2,n+1}^2x_{3,n+1}^2}{x_{13}^2}F_{n+1}\\
    =\,&-\frac14\log x_{12}^2\log x_{23}^2\times\lim_{\substack{x_{12}^2,x_{23}^2\to0\\x_{n+1}\to x_2}}\frac{x_{12}^2x_{23}^2x_{1,n+1}^2x_{2,n+1}^2x_{3,n+1}^2}{x_{13}^2}F_{n+1}.
\end{align*}
Therefore, we obtain the cusp limit constraint on $F_n$:
\begin{equation}\label{eq:ruleRational}
    \lim_{\substack{x_{12}^2,x_{23}^2\to0\\x_{n+1}\to x_2}}\frac{x_{12}^2x_{23}^2x_{1,n+1}^2x_{2,n+1}^2x_{3,n+1}^2}{x_{13}^2}F_{n+1} = 2\lim_{x_{12}^2,x_{23}^2\to0}x_{12}^2x_{23}^2F_n,
\end{equation}
where we have used $\Gamma_{\rm cusp}^{{\rm adj.}}|_{a^1}=2$, which could be obtained either from eq.(4.25) of~\cite{Alday:2010zy} ($-\frac14\Gamma_{\rm cusp}^{{\rm adj.}}|_{a^1}\frac{g_{\rm YM}^2N_c}{4\pi^2}=-\frac{g_{\rm YM}^2N_c}{8\pi^2}$), or by plugging in the $n=5$ result~\cite{Eden:2011we}:
\begin{equation}\label{eq:n56result}
    F_5=\frac1{\prod_{1\leq i<j\leq5}x_{ij}^2},\quad F_6=\frac{\frac1{48}\sum_{\sigma\in S_6}x_{\sigma_1\sigma_2}^2x_{\sigma_3\sigma_4}^2x_{\sigma_5\sigma_6}^2}{\prod_{1\leq i<j\leq6}x_{ij}^2}.
\end{equation}

\section{A New Graphical Rule}\label{sec:rule}

Since the rational function $F_n(x_1,\cdots,x_n)$ is $S_n$-symmetric and has conformal weight 4 with respect to each $x_i$, it can be written as a linear combination $F_n=\sum_{i=1}^{\mathcal N_n}c_{i}^{(n)}f_{i}^{(n)}$ of ``$f$-graphs''~\cite{Eden:2011we}. Here, $\mathcal N_n$ denotes the number of $n$-point $f$-graphs; the precise values are recorded in Table~\ref{tab:counting}. Each $f$-graph (after performing \texttt{Expand[]} in Mathematica) represents a permutation-invariant sum of terms, normalized such that the coefficient of any term equals 1. Denominators are represented by solid lines while numerators are represented by dashed lines. For example, there is a unique $f$-graph at $n=6$:
\begin{equation}
    F_6=f_{1}^{(6)},\quad f_{1}^{(6)}=\includegraphics[scale=0.2,align=c]{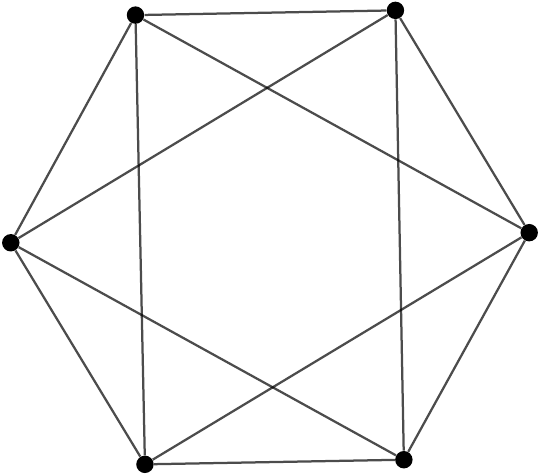}=\frac{\frac1{48}\sum_{\sigma\in S_6}x_{\sigma_1\sigma_2}^2x_{\sigma_3\sigma_4}^2x_{\sigma_5\sigma_6}^2}{\prod_{1\leq i<j\leq6}x_{ij}^2}=\frac{x_{12}^2x_{34}^2x_{56}^2+\text{perm.}}{\prod_{1\leq i<j\leq6}x_{ij}^2}.
\end{equation}
In other words, $f_{i}^{(n)}$ represents a sum of $n!/|f_{i}^{(n)}|$ terms, where $|f_{i}^{(n)}|$ is the graph symmetry factor\footnote{The symmetry factor of an $f$-graph can be computed using \texttt{GroupOrder@GraphAutomorphismGroup[]} in Mathematica; see the ancillary file \texttt{demo.nb} for practical tricks to handle $f$-graphs with numerators.}; for example, $|f_{1}^{(6)}|=48$.

\begin{table}[h]
    \centering
    \begin{tabular}{||c|ccc||}
        \textbf{$n$} & \textbf{all $f$-graphs ($\mathcal N_n$)} & \textbf{planar $f$-graphs ($N_n$)} & \textbf{plane-embedded $f$-graphs}\\
        \hline
       5 & 1 & 0 & 0 \\
       6 & 1 & 1 & 1 \\
       7 & 4 & 1 & 1 \\
       8 & 32 & 3 & 3 \\
       9 & 930 & 7 & 7 \\
       10 & 189341 & 36 & 36 \\
       11 & $\cdots$ & 220 & 220 \\
       12 & & 2707 & 2709 \\
       13 & & 42979 & 43017 \\
       14 & & 898353 & 900145 \\
       15 & & 22024902 & 22097042 \\
       16 & & 619981403 & $\cdots$
    \end{tabular}
    \caption{Number of $f$-graphs, which determines the size of the ansatz in the graphical bootstrap. All $f$-graphs including nonplanar ones can be generated using \texttt{nauty}~\cite{MCKAY201494}, and the plane-embedded $f$-graphs can be generated using \texttt{plantri}~\cite{Brinkmann2007FastGO}. The third column is used in~\cite{Bourjaily:2016evz} in order to apply the square and pentagon rules, but it will be of no use to us because our graphical rules are oblivious to embeddings.}
    \label{tab:counting}
\end{table}

The cusp limit constraint eq.~\eqref{eq:ruleRational} translates into a nice graphical rule that relates $F_n$ and $F_{n-1}$. On the right-hand side, in order to survive the limit, a term in $F_{n-1}$ must contain denominators $1/(x_{12}^2x_{23}^2)$; these terms are grouped into (solid-line-)cusp-highlighted $f$-graphs $f_{j,\alpha}^{\vee(n-1)}$, each representing a sum of $2(n-4)!/|f_{j,\alpha}^{\vee(n-1)}|$ terms with unit coefficients related by permuting $\{1,3\}$ and $\{4,5,\cdots,n-1\}$\footnote{The symmetry factor of highlighted $f$-graphs can also be computed in Mathematica after introducing some auxiliary vertices and edges to differentiate the highlighted pieces from the rest of the $f$-graph; see \texttt{demo.nb} for details.}. In general, there are $\mathcal V_{j}^{(n-1)}$ inequivalent ways to highlight a cusp in $f_{j}^{(n-1)}$. For example, we have $\mathcal V_{1}^{(6)}=2$, and the symmetry factors $|f_{1,1}^{\vee(6)}|=2$, $|f_{1,2}^{\vee(6)}|=4$:
\begin{align*}
    f_{1,1}^{\vee(6)}&=\includegraphics[scale=0.2,align=c]{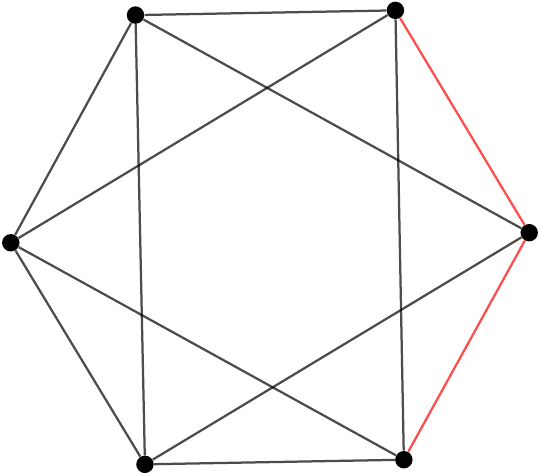}=\frac1{{\color{red}x_{12}^2x_{23}^2}x_{34}^2x_{45}^2x_{56}^2x_{16}^2x_{13}^2x_{35}^2x_{15}^2x_{24}^2x_{46}^2x_{26}^2}+{\rm perm}(4,5,6);\\
    f_{1,2}^{\vee(6)}&=\includegraphics[scale=0.2,align=c]{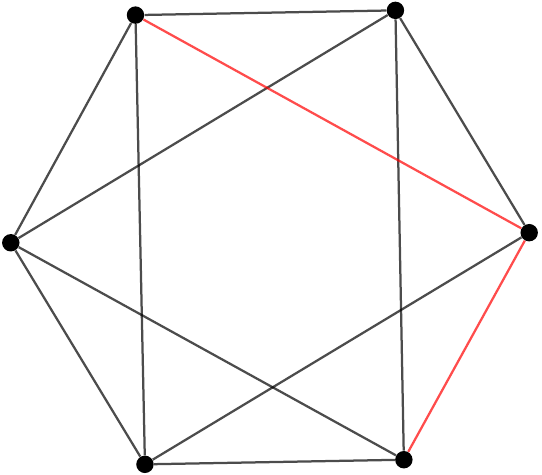}=\frac1{{\color{red}x_{12}^2}x_{24}^2x_{34}^2x_{35}^2x_{56}^2x_{16}^2x_{14}^2x_{45}^2x_{15}^2{\color{red}x_{23}^2}x_{36}^2x_{26}^2}+{\rm cyc}(4,5,6).
\end{align*}
With these definitions, we have
\begin{equation}
    \lim_{x_{12}^2,x_{23}^2\to0}x_{12}^2x_{23}^2f_{j}^{(n-1)}=x_{12}^2x_{23}^2\sum_{\alpha=1}^{\mathcal V_{j}^{(n-1)}}f_{j,\alpha}^{\vee(n-1)}.
\end{equation}
Similarly, on the left-hand side, in order to survive the limit, a term in $F_n$ must contain denominators $1/(x_{12}^2x_{23}^2x_{1n}^2x_{2n}^2x_{3n}^2)$; these terms are grouped into (solid-line-)double-triangle-highlighted $f$-graphs $f_{i,\mu}^{\diamond(n)}$, each representing a sum of $4(n-4)!/|f_{i,\mu}^{\diamond(n)}|$ terms with unit coefficients related by permuting $\{2,n\}$, $\{1,3\}$, and $\{4,5,\cdots,n-1\}$. In general, there are $\mathcal D_{i}^{(n)}$ inequivalent ways to highlight a double-triangle in $f_{i}^{(n)}$. For example, $\mathcal D_{1}^{(6)}=1$, and the symmetry factor $|f_{1,1}^{\diamond(6)}|=4$:
\begin{align*}
    f_{1,1}^{\diamond(6)}&=\includegraphics[scale=0.3,align=c]{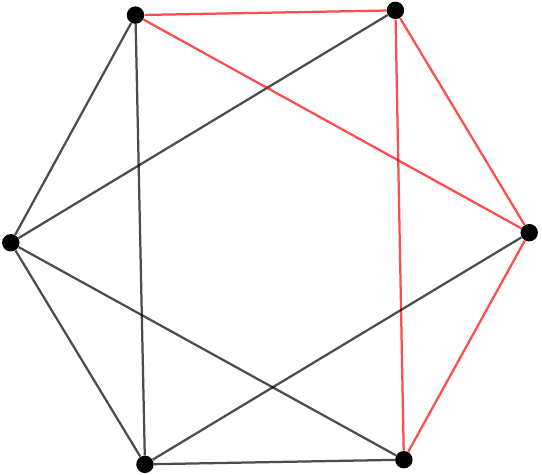}=\frac1{{\color{red}x_{12}^2x_{26}^2x_{36}^2}x_{34}^2x_{45}^2x_{15}^2{\color{red}x_{16}^2}x_{46}^2x_{14}^2{\color{red}x_{23}^2}x_{35}^2x_{25}^2}+(2\leftrightarrow6).
\end{align*}
Note that some $f$-graphs (first appearing at $n=12$) do not have double-triangle subgraphs and the corresponding $\mathcal D_i^{(n)}=0$. With these definitions,
\begin{equation}
    \lim_{\substack{x_{12}^2,x_{23}^2\to0\\x_n\to x_2}}\frac{x_{12}^2x_{23}^2x_{1n}^2x_{2n}^2x_{3n}^2}{x_{13}^2}f_{i}^{(n)}=x_{12}^2x_{23}^2\lim_{x_n\to x_2}\frac{x_{1n}^2x_{2n}^2x_{3n}^2}{x_{13}^2}\sum_{\mu=1}^{\mathcal D_{i}^{(n)}}f_{i,\mu}^{\diamond(n)}.
\end{equation}
The limit $x_n\to x_2$ is realized graphically by the operation $\mathcal P:f^{\diamond(n)}\mapsto f^{\vee(n-1)}$ that replaces the highlighted double-triangle with a highlighted cusp:
\begin{gather}\label{eq:grule}
    \lim_{x_n\to x_2}\frac{x_{1n}^2x_{2n}^2x_{3n}^2}{x_{13}^2}f_{i,\mu}^{\diamond(n)}=2\frac{|\mathcal P(f_{i,\mu}^{\diamond(n)})|}{|f_{i,\mu}^{\diamond(n)}|}\mathcal P(f_{i,\mu}^{\diamond(n)}),\quad\mathcal P:\ \includegraphics[scale=0.5,align=c]{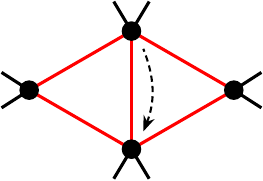} \implies \includegraphics[scale=0.5,align=c]{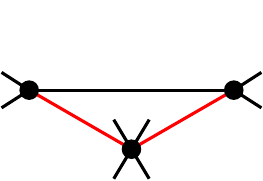}\ .
\end{gather}
The factor $2\frac{|\mathcal P(f_{i,\mu}^{\diamond(n)})|}{|f_{i,\mu}^{\diamond(n)}|}$ is needed so that $2\frac{|\mathcal P(f_{i,\mu}^{\diamond(n)})|}{|f_{i,\mu}^{\diamond(n)}|}\mathcal P(f_{i,\mu}^{\diamond(n)})$ correctly represents $4(n-4)!/|f_{i,\mu}^{\diamond(n)}|$ terms, if terms with non-unit coefficients are counted multiple times. Note that apart from the highlighted cusp, $\mathcal P$ introduces an additional denominator that could cancel against existing numerators. For example,
\begin{gather}
    \mathcal P(f_{1,1}^{\diamond(6)})=f_{1,1}^{\vee(5)} = \includegraphics[scale=0.3,align=c]{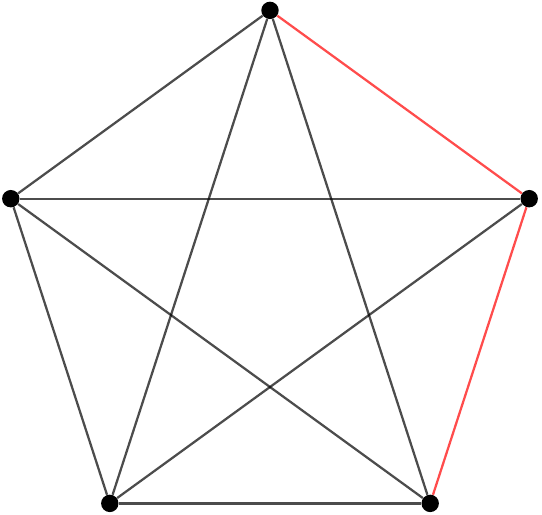}=\frac1{{\color{red}x_{12}^2x_{23}^2}x_{34}^2x_{45}^2x_{15}^2x_{13}^2x_{24}^2x_{35}^2x_{14}^2x_{25}^2},\\
    f_{1,1}^{\diamond(7)} = \includegraphics[scale=0.3,align=c]{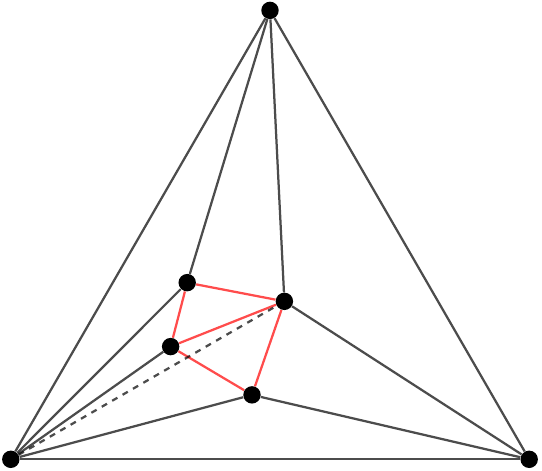}\xrightarrow{\mathcal P}\includegraphics[scale=0.3,align=c]{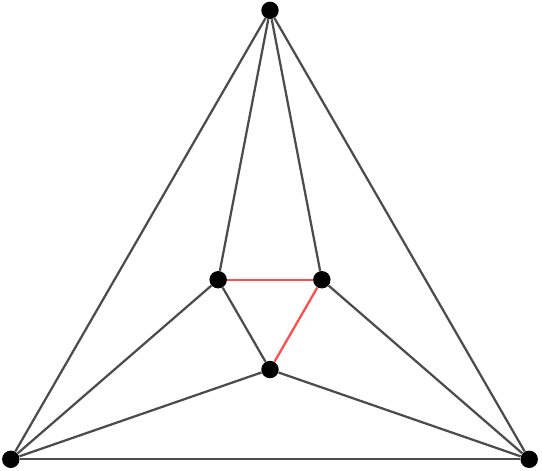}=f_{1,1}^{\vee(6)},\label{eq:pinch761}\\
    f_{1,2}^{\diamond(7)}=\includegraphics[scale=0.3,align=c]{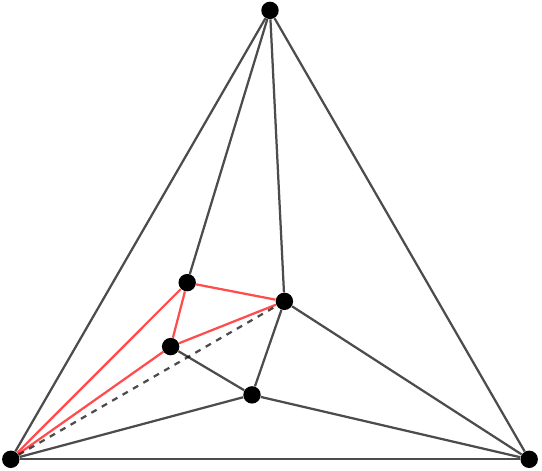}\xrightarrow{\mathcal P}\includegraphics[scale=0.3,align=c]{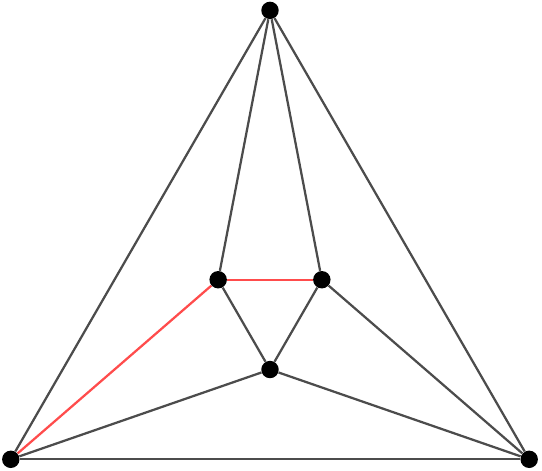}=f_{1,2}^{\vee(6)}.\label{eq:pinch762}
\end{gather}
Another phenomenon worth mentioning is that since $\mathcal P$ pinches two vertices, it may create a double pole in the resulting $f^{\vee(n-1)}$. This first appears at $n=7$:
\begin{equation}
    \includegraphics[scale=0.3,align=c]{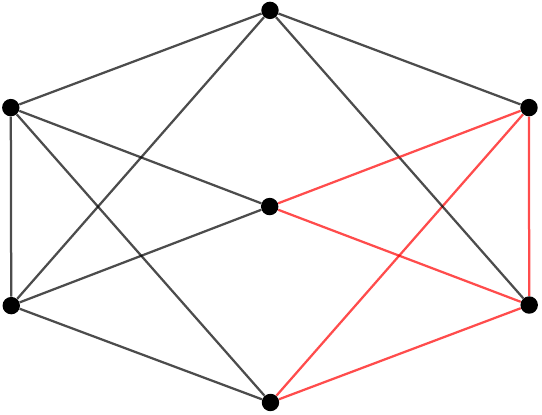} \xrightarrow{\mathcal P} \includegraphics[scale=0.3,align=c]{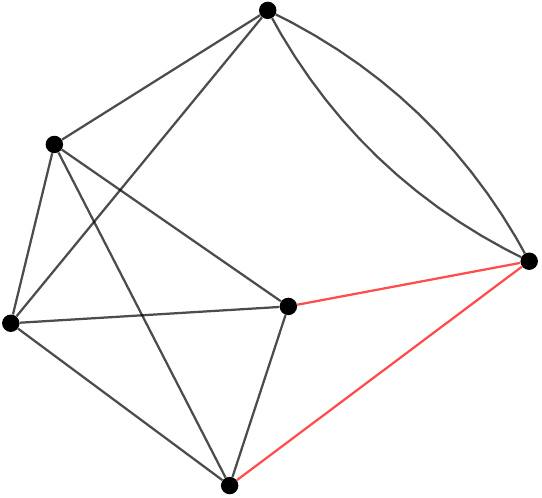} \xleftarrow{\mathcal P} \includegraphics[scale=0.3,align=c]{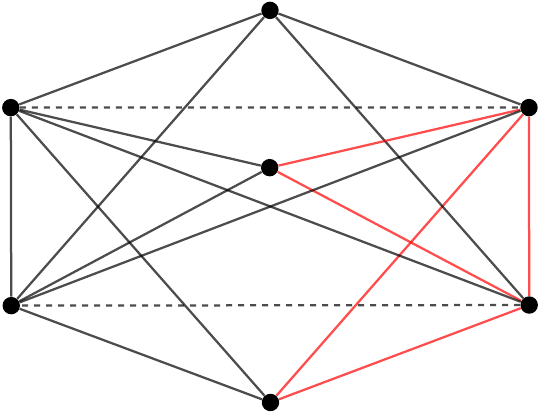}
\end{equation}
Such cusp-highlighted $f^{\vee(n-1)}$-graphs with a double pole do not arise from highlighting an $(n-1)$-point $f$-graph. For this reason, we denote them by $f_{0,\alpha}^{\vee(n-1)}$, where the subscript $_0$ reminds us they contribute to $F_{n-1}$ with coefficient 0; we stress that the notation $f_0^{(n-1)}$ does not make sense as it does not represent an $f$-graph. Now, putting the pieces together, we see that eq.\eqref{eq:ruleRational} implies
\begin{equation}
    \sum_{i=1}^{\mathcal N_n}\sum_{\mu=1}^{\mathcal D_{i}^{(n)}}\frac{|\mathcal P(f_{i,\mu}^{\diamond(n)})|}{|f_{i,\mu}^{\diamond(n)}|}c_{i}^{(n)}\mathcal P(f_{i,\mu}^{\diamond(n)})=\sum_{j=1}^{\mathcal N_{n-1}}\sum_{\alpha=1}^{\mathcal V_{j}^{(n-1)}}c_{j}^{(n-1)}f_{j,\alpha}^{\vee(n-1)}.
\end{equation}
Comparing the coefficients of each cusp-highlighted $f^{\vee(n-1)}$, we obtain a set of \emph{bootstrap equations} constraining the $f$-graph coefficients $\{c_{i}^{(n)},c_{j}^{(n-1)}\}$:
\begin{equation}\label{eq:pinchRule}
    \forall f_{j,\alpha}^{\vee(n-1)}:\quad\sum_{f_{i,\mu}^{\diamond(n)}} \frac{|f_{j,\alpha}^{\vee(n-1)}|}{|f_{i,\mu}^{\diamond(n)}|} c_{i}^{(n)}=c_{j}^{(n-1)}; \quad \text{in particular, }c_0^{(n-1)}\equiv0,
\end{equation}
where the sum runs over $f_{i,\mu}^{\diamond(n)}$ such that $\mathcal P(f_{i,\mu}^{\diamond(n)})=f_{j,\alpha}^{\vee(n-1)}$. Of course, not all of these equations are independent; for example, both eq.\eqref{eq:pinch761} and eq.\eqref{eq:pinch762} leads to the same equation $c_{1}^{(7)}=c_{1}^{(6)}$. 

For brevity, we will often denote $c_{i}^{(n)}\equiv c_i$ and $c_{i}^{(n-1)}\equiv b_i$ if the value of $n$ is clear from context; so we could also say that the $7\to6$ bootstrap equation is $c_1=b_1$.

Before ending this section, let us make a few comments. We distinguish two types of graphical rules in the literature: our new graphical rule derived above and the triangle rule (see appendix~\ref{app:triangle}) apply generally to all $f$-graphs, planar or non-planar, while the square rule and the pentagon rule~\cite{Bourjaily:2016evz} relies on the correlator/amplitude duality which only holds in the planar limit. As we will see in the next section, when we do restrict to the planar limit, our new graphical rule, by itself, has the same constraining power as the triangle, square, and pentagon rules combined, at least up to $n=14$.

Another point worth mentioning is that although we have derived the new graphical rule from the cusp limit of correlators, the graphical rule (in suitably weakened forms) lends itself to other interesting interpretations. For example, instead of viewing one vertex of the double triangle as a Lagrangian insertion, we can treat all $n$ vertices as external. Instead of taking $x_{12}^2,x_{23}^2\to0$, we can take all $x_{i,i+1}^2\to0$ (null $n$-gon limit). According to the correlator/amplitude duality, in the planar limit, the operation $\mathcal P$ can be interpreted as the \emph{soft limit} of the squared tree amplitudes\footnote{Had we taken a null $m$-gon limit with $m<n$ instead, we would find the soft limit of loop integrands of squared amplitudes.}:
${\cal F}_n \to 2 {\cal F}_{n{-}1}$ where ${\cal F}_n=\mathcal A_n^2$ stands for the squared tree amplitude. This simply follows from the soft limit of tree amplitudes $\mathcal A_n\to\mathcal A_{n-1}$, and it has been discussed very recently in~\cite{He:2024hbb} in the context of splitting function/energy correlators in the collinear limit. In fact, this was how we started this project: before we knew about the cusp limit and the graphic rule, we found that this tree-level soft limit alone suffices to fix all coefficients of the planar $f$-graphs up to $n=12$ (eight loops). Note that this is much more special than the graphical rule described above, since it is algebraic (not graphical) and we need to take $n$-gon lightlike limits. Starting at $n\geq13$ we really need the full-fledged graphical rule described in this section.

\section{Results}\label{sec:res}

\subsection{Planar integrand up to 11 loops}\label{sec:boots}

We are now ready to bootstrap the integrand of $G^{(\ell)}(u,v)$ or the Born-level correlator $F_{4+\ell}(x_1,\cdots,x_{4+\ell})$. Simply write down an ansatz $F_n=\sum_{i=1}^{\mathcal N_n}c_{i}^{(n)}f_{i}^{(n)}$ of $f$-graphs (Table \ref{tab:counting}) and impose the bootstrap equations on the coefficients. Importantly, since we have kept $N_c$ arbitrary, the cusp limit constraint and the graphical rule apply generally. However, let us now focus on the planar limit $N_c\to\infty$, where only planar $f$-graphs contribute for $n\geq6$~\cite{Eden:2012tu}\footnote{Note that $n=5$ (one loop) is an exception, where the unique $f$-graph is nonplanar.}. We will return to the nonplanar case in section~\ref{sec:nonplanar}.

\begin{figure}[H]
    \centering
    \includegraphics[width=0.98\textwidth]{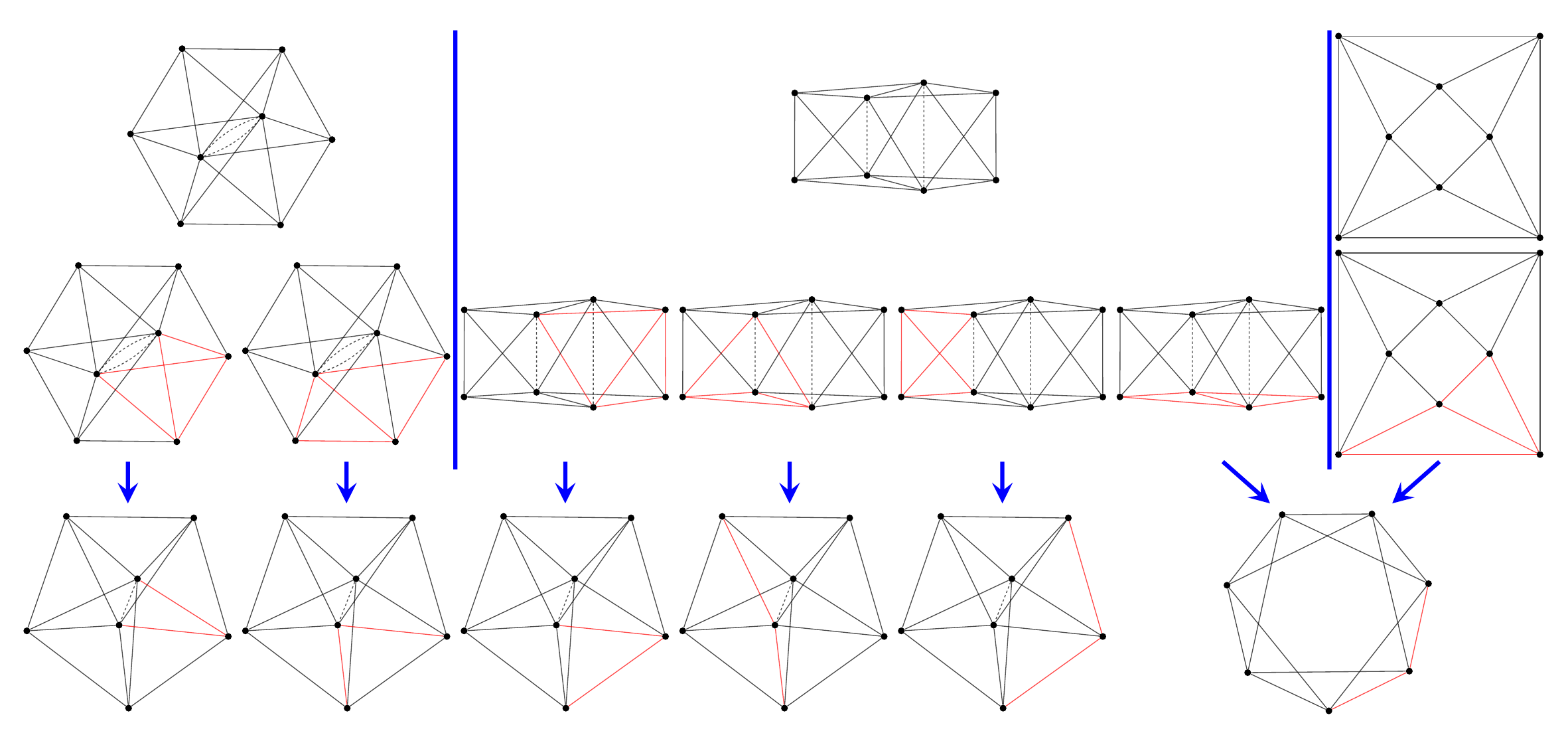}
    \caption{The graphical rule relating $F_8$ and $F_7$ in the planar limit. The first row shows all 3 planar $f$-graphs, which are ordered $f_{2}^{(8)},f_{3}^{(8)},f_{1}^{(8)}$ for convenience of plotting. The second row shows all double-triangle-highlighted $f$-graphs $f^{\diamond(8)}$. Performing the $\mathcal P$ operation yields the cusp-highlighted $f^{\vee(7)}$ in the third row, where the first five arise from the planar $f_{1}^{(7)}$ and the last one arise from a nonplanar $f$-graph $f_{3}^{(7)}$. In the planar limit, the nonplanar $f_{3}^{(7)}$ does not contribute to $F_7$ (i.e., $c_{3}^{(7)}=O(N_c^{-2})$), so the graphical rule leads to the $8\to7$ bootstrap equations $\{c_2=b_1,\ c_3=b_1,\ c_3+c_1=0\}$. In this case, there are no non-trivial symmetry factors in the bootstrap equations.}
    \label{fig:4lp}
\end{figure}

Figure~\ref{fig:4lp} demonstrates the $8\to7$ bootstrap equations in the planar limit, where the graphical rule imposes three equations\footnote{Note that our numbering of planar $f$-graphs is different from~\cite{Bourjaily:2016evz}. Our reason is that, for large enough $n$ (Table~\ref{tab:counting}), the set of planar $f$-graphs that we use is completely different from the set of plane-embedded $f$-graphs, so in general we cannot follow the numbering of~\cite{Bourjaily:2016evz} and may as well use a different numbering from the beginning.}:
\begin{equation}\label{eq:bootstrap8}
c_2=b_1,\quad c_3=b_1,\quad c_1+c_3=0. \end{equation}
Solving the bootstrap equations fixes $c_{1,2,3}$ and $b_1$ up to an overall normalization. Using the $7\to6$ and $6\to5$ bootstrap equations $c_{1}^{(7)}=c_{1}^{(6)}=c_{1}^{(5)}$, coefficients across different loop orders are fixed in terms of single normalization $c_{1}^{(5)}$ which is then fixed by our convention to be $c_{1}^{(5)}=1$ (eq.\eqref{eq:n56result}). It turns out that the $n\to(n-1)$ bootstrap equations uniquely fix $\{c_{i}^{(n)}\}$ up to $n=11$.

\begin{figure}[h]
    \centering
    $\left(\includegraphics[scale=0.2,align=c]{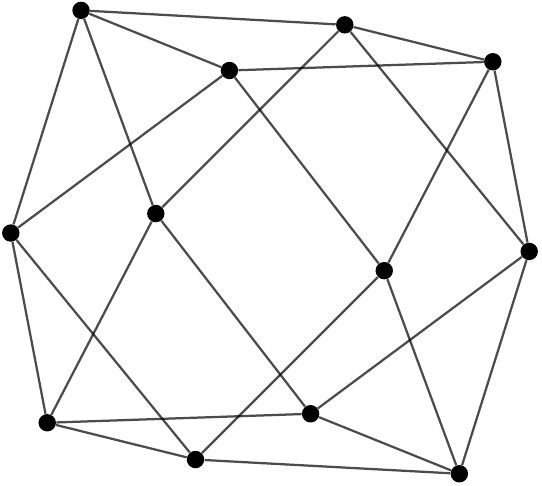}=\right)$\includegraphics[scale=0.2,align=c]{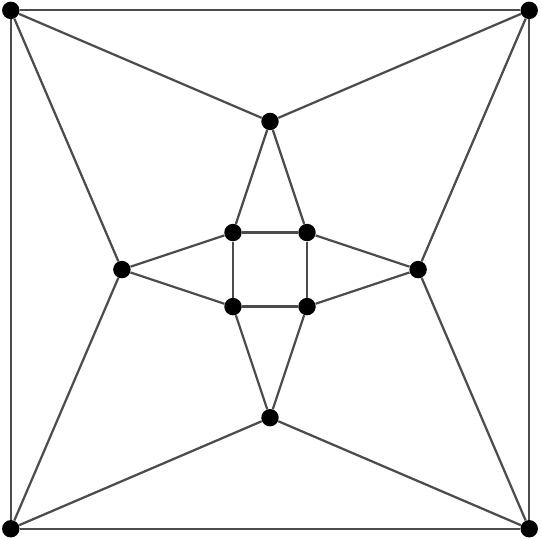}\hspace{2em}\includegraphics[scale=0.2,align=c]{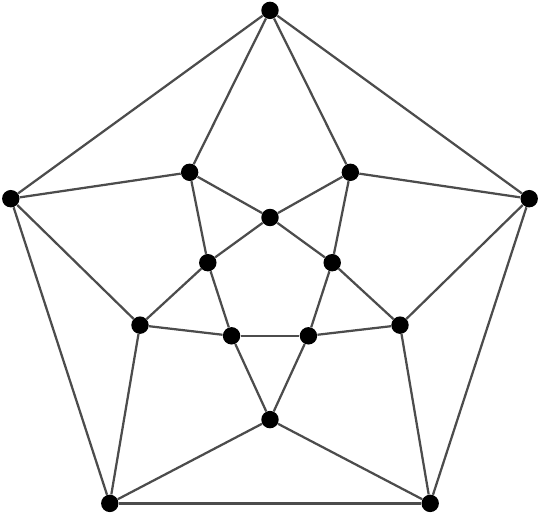}\hspace{2em}\includegraphics[scale=0.2,align=c]{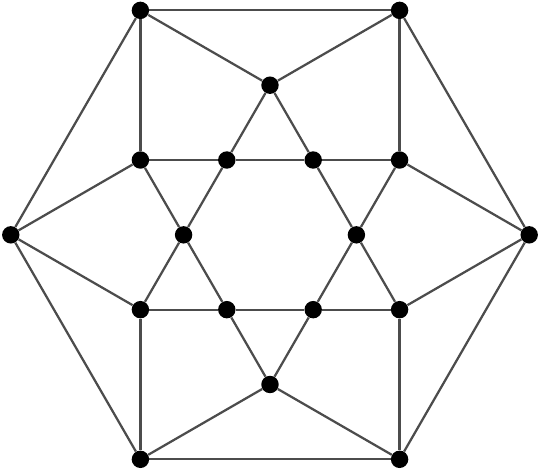}\hspace{2em}$\cdots$\\\vspace{1em}
    \includegraphics[scale=0.2,align=c]{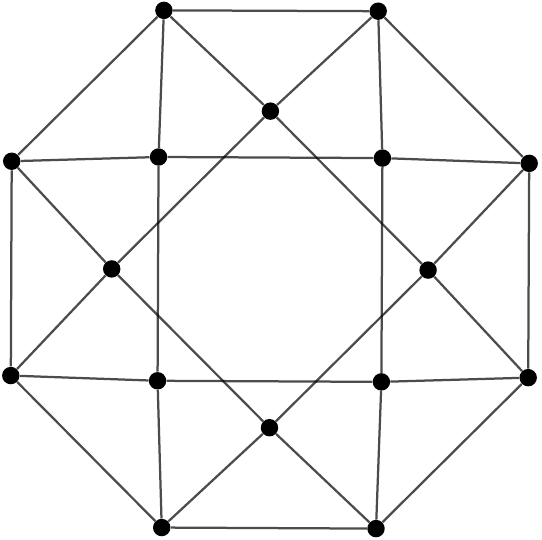}\hspace{2em}\includegraphics[scale=0.2,align=c]{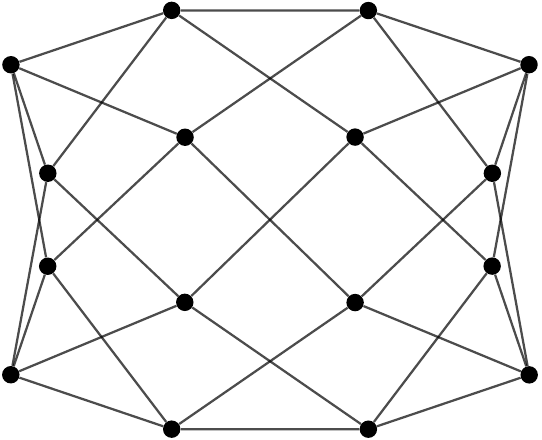}
    \caption{Planar $f$-graphs without double-triangle subgraphs. The first row demonstrates the family of ``chopped prisms'' that appear at $n=12,15,18,\cdots$, so named because the $n=3m$ chopped prism can be viewed as a $m$-prism with all $2m$ vertices chopped (see the different drawing for $n=12$). The second row displays the two $16$-point $f$-graphs invisible to the $16\to15$ bootstrap equations. We have checked by enumeration that there are only 4 planar $f$-graphs without double-triangle subgraphs for $n\leq16$, which are all shown above.}
    \label{fig:uds}
\end{figure}

At $n=12$ (eight loops), a complication arises because there is an $f$-graph without double-triangle subgraphs. The problem gets worse at higher points; in fact, one could easily construct an infinite family of $f$-graphs (``chopped prisms'') without double-triangle subgraphs (Figure~\ref{fig:uds}). These $n$-point $f$-graphs are ``invisible'' in the $n\to(n-1)$ bootstrap equations as their coefficients do not appear and hence are not constrained.
On the contrary, all $c_i^{(n-1)}$ coefficients will appear in the $n \to n-1$ bootstrap equations as any $f$-graph contains a cusp structure.

It came to us as such a surprise that, although the $12\to11$ bootstrap equations miss the 12-point chopped prism entirely, its coefficient is fixed by the $13\to12$ bootstrap equations! In fact, the entire set $\{c_{i}^{(12)}\}$ of 12-point coefficient is fixed (together with $\{c_{i}^{(13)}\}$) by the $13\to12$ bootstrap equations up to a single normalization. This phenomenon that \textbf{the $(n+1)\to n$ bootstrap equations fixes all visible planar coefficients $\{c_{i}^{(n)},c_{i}^{(n+1)}\}$ up to a single normalization}\footnote{This is similar to what was observed in the triangle rule constraints in~\cite{Bourjaily:2016evz}, where the $(n+1)\to n$ triangle rule equations impose certain consistency conditions on $\{c_i^{(n)}\}$. However, they are not strong enough to completely fix all of these coefficients.} is checked up to $n=14$. For example, the $9\to8$ bootstrap equations read
\begin{equation}\label{eq:bootstrap9}
    \begin{gathered}
        c_2=c_3=2c_2+c_5=b_1,\\
        c_4=c_5=c_6=b_2,\\
        c_5=c_6=c_7=b_3,\\
        c_1+c_3=c_2+c_5=c_2+c_6=c_2+c_7=c_3+c_7=0,
    \end{gathered}
\end{equation}
which not only fix $(c_1,\cdots,c_7)=b_1\times(-1,1,1,-1,-1,-1,-1)$, but also imposes the constraint $b_2=-b_1$ and $b_3=-b_1$. We see that the coefficients $c_{1}^{(8)},c_{2}^{(8)},c_{3}^{(8)}$ are determined by the $9\to8$ bootstrap equations eq.\eqref{eq:bootstrap9} up to an overall normalization, just as they are determined by the $8\to7$ bootstrap equations eq.\eqref{eq:bootstrap8}. We stress that this is conceptually different from the idea of recursion that determines $\{c_{i}^{(n)}\}$ given $\{c_{i}^{(n-1)}\}$; rather, we can take the $(n+1)\to n$ bootstrap equations and directly obtain $\{c_{i}^{(n)}\}$ without knowing $\{c_{i}^{(n-1)}\}$.

Using the $(n+1)\to n$ bootstrap equations, we obtain a unique solution of $\{c_{i}^{(n)}\}$ for all $n\leq14$, and the result agrees precisely with~\cite{Bourjaily:2016evz}. Compared to their method which uses 3 different graphical rules, only one graphical rule is needed in our method, which follows from the simple cusp limit constraint. Moreover, the entire computation can be done within 3 hours on a workstation with 32GB memory and 32 Intel Xeon CPU E5-2620 v4 @ 2.10GHz.

For $n=15$, we again have a single $f$-graph (15-point chopped prism) invisible to the $15\to14$ bootstrap equations. Although we strongly suspect that the $16\to15$ bootstrap equations will completely fix $\{c_{i}^{(15)}\}$, that computation would take two months to complete by a rough estimate. For the time being, we use the triangle rule\footnote{See~\cite{Bourjaily:2016evz} for a nice discussion. However, some subtleties in the proof of the triangle rule were not addressed in that paper. We provide a more rigorous argument in appendix~\ref{app:triangle}.} relating $F_{15}$ and $F_{14}$ to fix its coefficient. Additionally, the triangle rule serves as a highly non-trivial consistency check of our results. The entire computation takes around a week on the aforementioned platform. 

Let us briefly comment on the relation between our graphical rule and the triangle rule. On one hand, the triangle rule has the advantage that no planar $f$-graph is invisible to the triangle rule\footnote{That all planar $f$-graphs must have triangle subgraphs can be shown by a simple graph-theoretical argument.}. On the other hand, overall, the triangle rule is much weaker than our graphical rule. This is because the triangle rule compares\footnote{We use the figurative notation $f^\triangle,f^|,f^\bullet$ to denote the triangle-, edge-, and vertex-highlighted $f$-graphs needed in the description of the triangle rule. For details, see~\cite{Bourjaily:2016evz} where the shrinking operations $\mathcal T,\mathcal E$ are also defined.} $\mathcal T(f^{\triangle(n)})=f^{\bullet(n-2)}$ and $\mathcal E(f^{|(n-1)})=f^{\bullet(n-2)}$ which yields roughly $(n-2)N_{n-2}$ equations, while our graphical rule compares $\mathcal P(f^{\diamond(n)})=f^{\vee(n-1)}$ and $f^{\vee(n-1)}$ which yields roughly $(n-1)^3N_{n-1}$ equations, significantly more than the triangle rule. In a sense, the two rules are complementary to each other, capturing the physics of the correlator in different spacetime signatures: our graphical rule arises from the double-log divergence of the cusp limit in Minkowskian signature, while the triangle rule arises from the single-log divergence of the OPE limit in Euclidean signature. 

\begin{table}[h]
    \centering
    \begin{equation*}
        \begin{array}{||c||c|c|c|c|c|c|c|c|c|c|c|c|c|c|c|c||}
        n & 14 & \pm 5 & 4 & \pm 3 & \pm \frac{11}{4} &\pm \frac{5}{2} & \pm \frac{9}{4} & \pm 2 & \pm \frac{7}{4} & \pm \frac{3}{2} & \pm \frac{5}{4} & \pm 1 & \pm \frac{3}{4} & \pm \frac{1}{2} & \pm \frac{1}{4} & 0 \\
        \hline
        5 &&&&&&&&&&&&1&&&& \\
        6 &&&&&&&&&&&&1&&&& \\
        7 &&&&&&&&&&&&1&&&& \\
        8 &&&&&&&&&&&&3&&&& \\
        9 &&&&&&&&&&&&7&&&& \\
        10 &&&&&&&&1&&&&25&&&&10 \\
        11 &&&&&&&&1&&&&126&&&&93 \\
        12 &&1&&&&&&9&&3&&906&&141&&1647 \\
        13 &&1&&&&&&54&&22&&7919&&2529&&32454 \\
        14 &1&9&1&&&4&4&490&5&329&18&78949&559&50633&5431&761920 \\
        15 &1&57&8&6&2&45&54&4812&90&4264&480&843375&19664&892668&282736&19976640
        \end{array}
    \end{equation*}
    \caption{Distribution of planar coefficients $\{c_{i}^{(n)}\}_{i=1}^{N_n}$ up to 15 points. Note that the majority of coefficients are 0 (roughly 90\% at $n=15$). Also, it looks like there is always a unique $f$-graph with the largest absolute coefficient, which happens to coincide with the first few Catalan numbers~\cite{Bourjaily:2016evz}. We will discuss this conjecture in section~\ref{sec:catalan}.}
    \label{tab:stats}
\end{table}

Another observation is that it is surprisingly easy to solve the bootstrap equations. First, the majority of equations are the simplest possible ones, of the form $c_i=b_j$ (or $c_i=0$). Moreover, if $c_{i_1}$ appears in the simplest equations, substituting the solution into the next-to-simplest equations $c_{i_1}+c_{i_2}=b_j$ reduces it to a simplest equation. Iterating this process, we can solve more and more variables without performing Gauss elimination on dense matrices. The nice surprise is that, at least up to $n=15$, the $n\to(n-1)$ bootstrap equations can be completely solved in this way; dense blocks \emph{never} appear. As a result, the bootstrap equations can be very efficiently solved in terms of sparse matrix operations.

Regarding the results of $F_{\leq15}$, we find that the statistics of coefficients follow a nice pattern (see Table~\ref{tab:stats}). Firstly, new coefficients $\pm 3$, $\pm \frac{11}{4}$ appear in the $n=15$ case, as opposed to $n \leq 14$, where new coefficients only appear at even loops. Secondly, for $n\leq11$ the coefficients are all integers ($\pm1$, $\pm2$), while starting at $n=12$ the smallest interval gets halved every two loops: $1/2$ for $n=12,13$ and $1/4$ for $n=14, 15$. The reason for this lies in the symmetry factors $|f_{j,\alpha}^{\vee(n-1)}|/|f_{i,\mu}^{\diamond(n)}|$ in the bootstrap equations.
For some reasons (which can be analyzed case by case for all graphs from $n=11$ to $n=15$), this halves the smallest interval for coefficients at even $n$ but not odd $n$. It would be interesting to analyze the combinatorial structure of the bootstrap equations to see if this continues to higher loops. Thirdly, the dominance of vanishing coefficients keeps growing with $n$: $60.8\%,75.5\%,84.8\%,90.7\%$ for $n=12,13,14,15$, which strongly suggests that some unknown selection rule is at work. However, at present, we have little understanding why these $f$-graphs should not appear in the final result. If we could somehow eliminate these unused $f$-graphs from the beginning, it would greatly increase the efficiency of the bootstrap.

As we have mentioned, upon taking various lightlike limits, the correlator integrand $F_n$ contains (squared) superamplitudes in planar ${\cal N}=4$ SYM of many loops and legs. The most immediate consequence of our new result is the $11$-loop integrand of the four-point amplitude, which can be extracted by finding all 4-cycles of the $f$-graphs.
It becomes more nontrivial to extract higher-point amplitudes~\cite{Heslop:2018zut}, but we can very easily extract higher-point squared (tree) amplitudes $\mathcal F_n$ by taking $n$-gon lightlike limits. As proposed very recently~\cite{He:2024hbb}, the rational DCI function $\mathcal F_n$ with only simple local poles immediately gives the integrand for the leading-order $(n{-}3)$-point energy correlators in the collinear limit. In this sense, our results give the energy correlator integrand up to $12$ points. Finally, we note that anti-prisms correspond to the pole-free constant in ${\cal F}_n$, which only exists for even $n$ and takes values $1, -1, 2, -5, 14, -42$ for $n=6,8,10,12,14, 16$ (see the next subsection). It would be interesting to understand these constants directly from the square of tree amplitudes.

\subsection{Local analysis: anti-prisms and the Catalan conjecture}\label{sec:catalan}

Instead of computing the complete set of bootstrap equations, it is often worthwhile to search for local subsystems of bootstrap equations. By ``local'', we mean the subsystem only involves a small (compared to $N_n$) amount of $f$-graphs and hence much more easily solved. In this subsection, we demonstrate the power of such local analysis by disproving the Catalan conjecture for $n=16$ without knowing the complete $16\to15$ bootstrap equations.

\begin{figure}
    \centering
    \begin{equation*}
        \begin{array}{cccc}
        \includegraphics[scale=0.3,align=c]{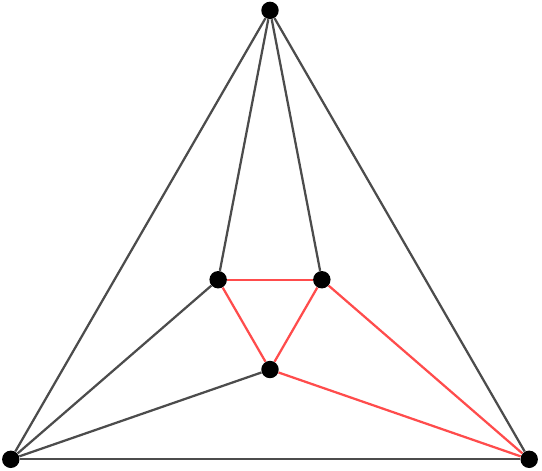}\quad & \includegraphics[scale=0.3,align=c]{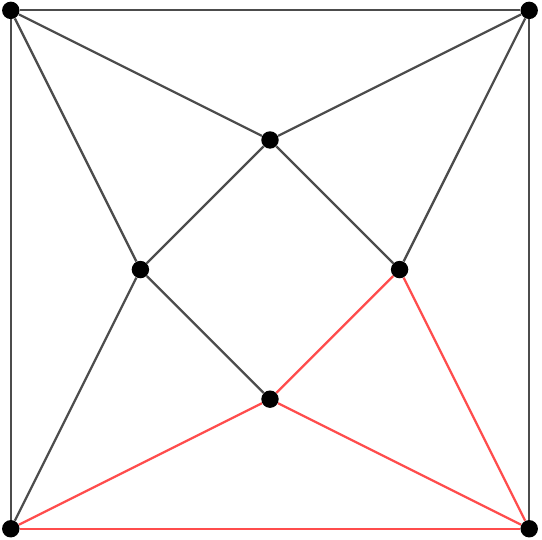}\quad & \includegraphics[scale=0.3,align=c]{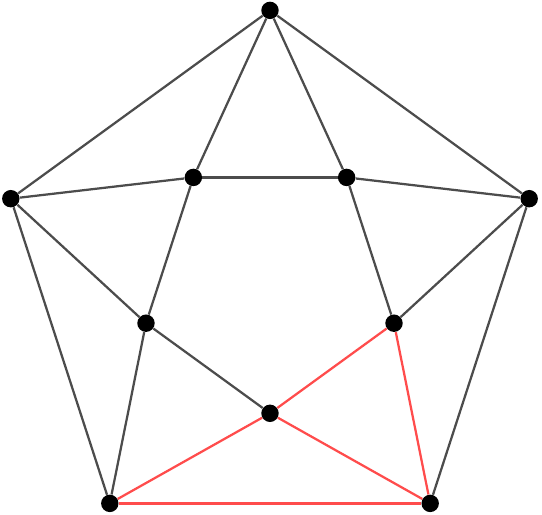}\quad & \includegraphics[scale=0.3,align=c]{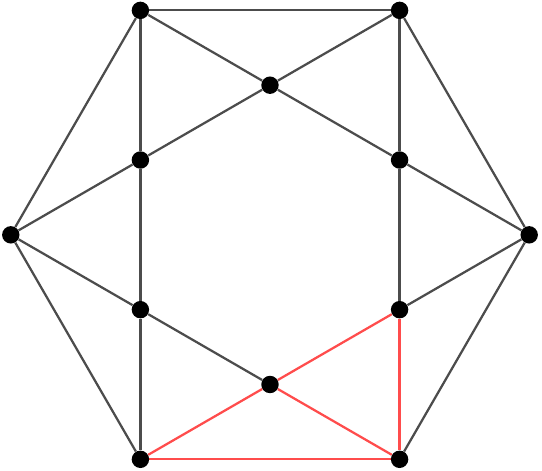} \\
        \includegraphics[scale=0.3,align=c]{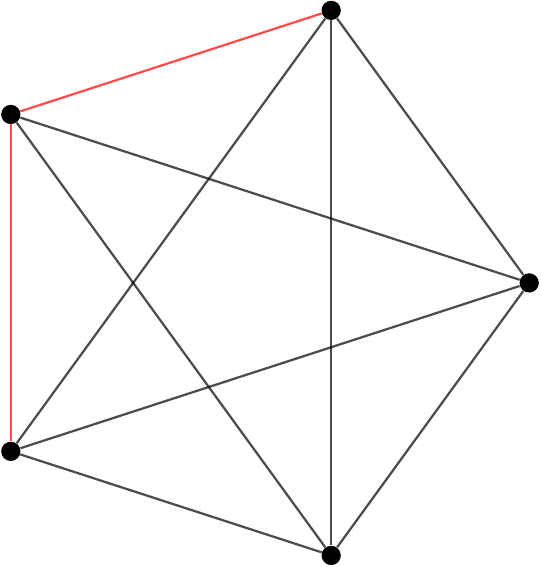}\quad & \includegraphics[scale=0.3,align=c]{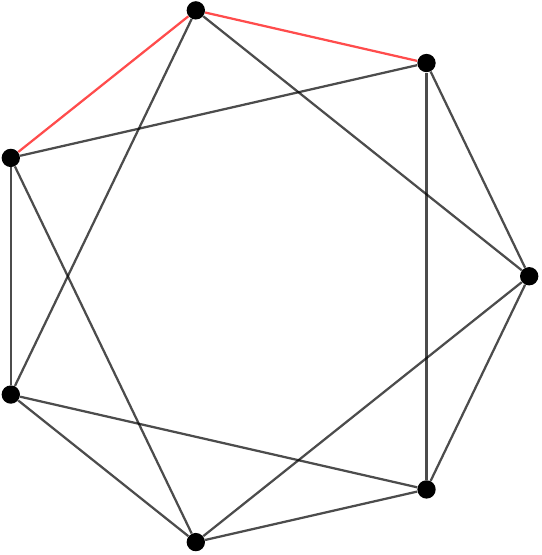}\quad & \includegraphics[scale=0.3,align=c]{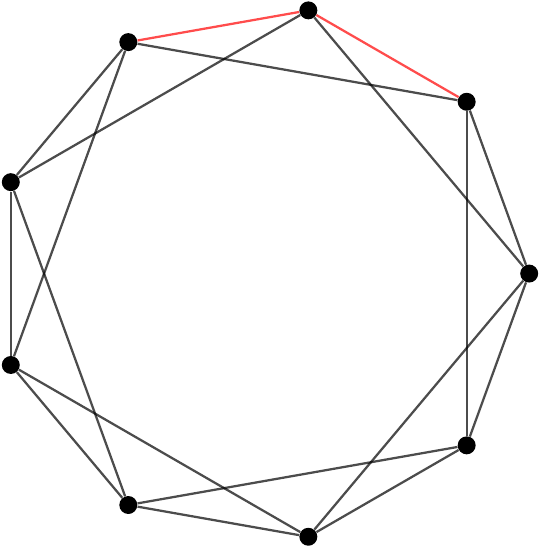}\quad & \includegraphics[scale=0.3,align=c]{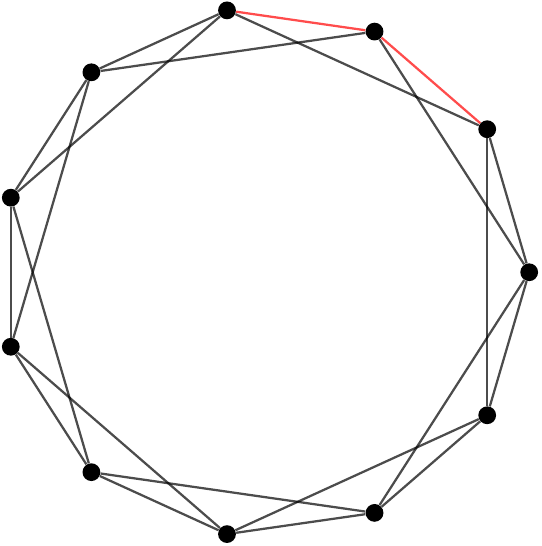}
        \end{array}
    \end{equation*}
    \caption{Anti-prisms and $\mathcal P$-anti-prisms. The first row shows the double-triangle-highlighted anti-prisms $f_{\rm anti}^{\diamond(n)}$ with $n=6,8,10,12$. Since there is a unique way to highlight a double-triangle in anti-prisms, we omit the second subscript. The second row shows the result of pinch $\mathcal P(f_{\rm anti}^{\diamond(n)})=f_{\xi,1}^{\vee(n-1)}$, where the subscript $\xi$ reminds us that $f_\xi^{(n-1)}=1/\xi_{n-1}+\text{perm.}$, where $\xi_n:=\prod_{i=1}^nx_{i,i+1}^2x_{i,i+2}^2$.}
    \label{fig:anti}
\end{figure}

Let us first recall the Catalan conjecture. In~\cite{Bourjaily:2016evz}, a curious pattern was observed that the largest absolute coefficient in $F_n$ seems to coincide with the first few Catalan numbers. For even $n$, the largest coefficient corresponds to the anti-prism $f$-graphs (Figure~\ref{fig:anti}). Although we have not obtained the complete $16\to15$ bootstrap equations, our graphical rule does enable us to study some aspects of $F_{16}$. In particular, we will show that the above pattern continues to hold at $n=16$: $c_{\rm anti}^{(16)}=-42$.

By symmetry, there is always a unique way to highlight a double-triangle in an anti-prism, and $f_{\rm anti}^{\diamond(n)}$ pinches to the nonplanar $f_{\xi,1}^{\vee(n-1)}$. In other words, $c_{\rm anti}^{(n)}$ appears in the $n\to(n-1)$ bootstrap equations only once. Therefore, we compute $c_{\rm anti}^{(n)}$ in two steps: (1) construct the $f_{\xi,1}^{\vee(n-1)}$ bootstrap equation, and (2) compute the other $c_i^{(n)}$ appearing in the equation independently. Then, we will be able to solve for $c_{\rm anti}^{(n)}$ and test the Catalan conjecture.

In order to construct the equation, we need to find all $f$-graphs that could pinch to $f_{\xi,1}^{\vee(n-1)}$, according to eq.\eqref{eq:pinchRule}. For $n=6$, no other $f$-graphs pinch to $f_{\xi,1}^{\vee(5)}$, and $c_{\rm anti}^{(6)}=c_1^{(5)}=1$. For $n\geq8$, there are other $f$-graphs that pinch to $f_{\xi,1}^{\vee(n-1)}$, which we will find by ``unpinching'' $f_{\xi,1}^{\vee(n-1)}$.

\begin{figure}[h]
    \centering
    $f_{\rm anti}^{\diamond(10)}=$\includegraphics[scale=0.8,align=c]{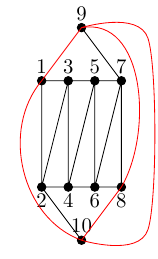}$\xrightarrow{\mathcal P}$\includegraphics[scale=0.8,align=c]{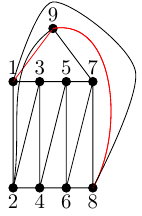}$\xrightarrow{\text{add vertex}}$\includegraphics[scale=0.8,align=c]{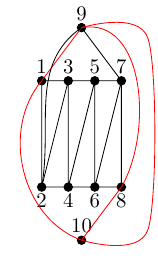}$\xrightarrow{\text{adjust weights}}\{f_{i,\mu}^{\diamond(10)}\}$
    \caption{Example of ``unpinching'' with $n=10$. On the left is a different drawing of the $\mathcal P(f_{\rm anti}^{\diamond(10)})=f_{\xi,1}^{\vee(9)}$. ``Unpinching'' consists of two steps: (1) expanding the cusp into a double-triangle which changes the conformal weight of 9 to five and introduces a new vertex 10 with conformal weight three; (2) adjusting the conformal weights of 9 and 10 without changing the conformal weights of other vertices. Note that the edge $2\bullet\!\!-\!\!\bullet9$ breaks planarity.}
    \label{fig:ant910}
\end{figure}

Let us illustrate the ``unpinching'' procedure with an $n=10$ example (Figure~\ref{fig:ant910}). We label the vertices temporarily in order to refer to them individually. ``Unpinching'' consists of first expanding the cusp into a double-triangle and then adjusting the conformal weights:
\begin{align*}
    \frac1{\xi_9}\xrightarrow{\text{add vertex}}&\frac1{\xi_9}\frac{x_{1,8}^2}{x_{1,10}^2x_{8,10}^2x_{9,10}^2}\\
    \xrightarrow{\text{adjust weights}}&\frac1{\xi_9}\frac{x_{1,8}^2}{x_{1,10}^2x_{8,10}^2x_{9,10}^2}\left(\frac{x_{3,10}^2}{x_{3,9}^2}\right)^{p_3}\left(\frac{x_{5,10}^2}{x_{5,9}^2}\right)^{p_5}\left(\frac{x_{2,9}^2}{x_{2,10}^2}\right)^{p_2}\left(\frac{x_{4,9}^2}{x_{4,10}^2}\right)^{p_4}\left(\frac{x_{6,9}^2}{x_{6,10}^2}\right)^{p_6}.
\end{align*}
Planarity forces $p_{3,5,4,6}\geq0$ and $p_2\geq1$  in order to cancel the $2\bullet\!\!-\!\!\bullet9$ denominator line. Forbidding double poles further imposes $p_{3,5,2,4,6}\leq1$. Finally, requiring vertices 9 and 10 to have conformal weight $4$ constrains $p_2+p_3+p_4+p_5+p_6=1$. In the end, the problem reduces to finding integer solutions to
\begin{equation}
    p_2=1,\quad p_3+p_5=p_4+p_6,\quad 0\leq p_{3,4,5,6}\leq1.
\end{equation}
The 6 solutions are
\begin{equation*}
    \begin{array}{ccc}
         (0,0,0,0)\sim f_{2}^{(10)}:\includegraphics[scale=0.6,align=c]{newimages/ant10ph.pdf}&(0,1,1,0)\sim f_{11}^{(10)}:\includegraphics[scale=0.6,align=c]{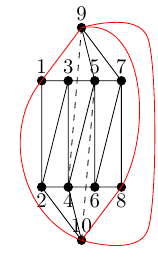}&(1,0,0,1)\sim f_{13}^{(10)}:\includegraphics[scale=0.6,align=c]{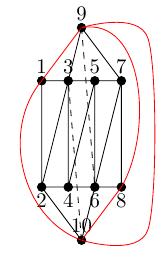}\\
         (1,1,1,1)\sim f_{30}^{(10)}:\includegraphics[scale=0.6,align=c]{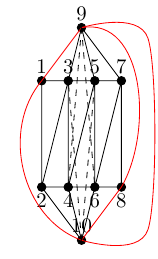}&(0,0,1,1)\sim f_{9}^{(10)}:\includegraphics[scale=0.6,align=c]{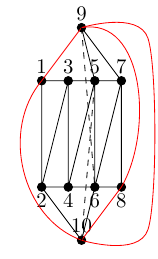}&(1,1,0,0)\sim f_{9}^{(10)}:\includegraphics[scale=0.6,align=c]{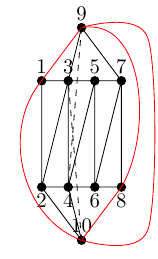}
    \end{array}
\end{equation*}
Note that $(0,0,1,1)$ and $(1,1,0,0)$ lead to the isomorphic $f$-graphs due to a 180$^\circ$-rotation symmetry. Therefore, the bootstrap equation involving $c_{\rm anti}^{(10)}=c_2^{(10)}$ reads
\begin{equation*}
    c_{2}+c_{11}+c_{13}+c_{30}+2c_{9}=0,
\end{equation*}
where the right hand side is zero because $f_\xi^{(9)}$ is nonplanar.

In general, the bootstrap equation involving $c_{\rm anti}^{(n)}$ can be found by solving
\begin{equation}
    \sum_{k=2}^{\frac{n}{2}-2} p_{2k-1}=\sum_{k=2}^{\frac{n}{2}-2} p_{2k},\quad 0\leq p_{2k-1},p_{2k}\leq1.
\end{equation}
A simple combinatoric argument shows that there are $\binom{n-6}{\frac{n}{2}-3}$ integer solutions. A closer look shows that these correspond to $\frac{\binom{n-6}{\frac{n}{2}-3}+2^{\frac{n}{2}-3}}{2}$ different $f$-graphs, $2^{\frac n2-3}$ of which are invariant under 180$^\circ$-rotation. As a result, the bootstrap equation reads
\begin{equation}\label{eq:anti}
    \sum_i \#_i c_i^{(n)}=0,\quad\text{where }\{\#_i\}\text{ contains }2^{\frac{n}{2}-3}\text{ 1's and }\frac{\binom{n-6}{\frac{n}{2}-3}-2^{\frac{n}{2}-3}}{2}\text{ 2's}.
\end{equation}

To determine the other $\{c_i^{(n)}\}$ appearing in the above equation, we use the similar pinch-unpinch method to ``locally'' generate bootstrap equations involving them. For $n=8,10,12,14$, we find that the subset of bootstrap equations involving only these $\frac{\binom{n-6}{\frac{n}{2}-3}+2^{\frac{n}{2}-3}}{2}=2,6,14,43$ coefficients form a closed system, which yields $c_{\rm anti}^{(8)}=-1$, $c_{\rm anti}^{(10)}=2$, $c_{\rm anti}^{(12)}=-5$, and $c_{\rm anti}^{(14)}=14$, confirming the Catalan conjecture.

At $n=16$, rather than locally generating all bootstrap equations involving the 142 coefficients,
we use the square rule\footnote{Strictly speaking, the square rule only applies for plane-embedded $f$-graphs. However, we have checked that all involved $f$-graphs have vertex-connectivity 4, which indicates that there is a unique plane embedding for each of them, so the use of the square rule is justified. Moreover, it can be shown that, for $f$-graphs with vertex-connectivity $\geq3$, the square rule is a special case of our graphical rule.}~\cite{Bourjaily:2016evz} to help speed up the calculation. The square rule relates a single 16-point $f$-graph (with a $\boxtimes$ subgraph) to a single 15-point $f$-graph and hence directly determines its coefficient. Of the 142 coefficients, 128 are directly related to 15-point coefficients in this way. For each of the remaining 13 coefficients (not including the anti-prism itself), we locally generate all bootstrap equations involving it and pick one that involves as less new $f$-graphs as possible. It turns out that the chosen bootstrap equations involve $(0,0,0,0,0,0,1,1,1,1,2,2,3)$ new $f$-graphs not in the list of 142, and all 11 new $f$-graphs are directly related to 15-point coefficients through the square rule. In summary, we have found enough bootstrap equations that relate $c_{\rm anti}^{(16)}$ to 152 other 16-point coefficients which are fixed by the square rule and our graphical rule. Solving this subsystem and plugging in the correct 15-point coefficients yields $c_{\rm anti}^{(16)}=-42$, confirming the Catalan conjecture. We include the list of these $f$-graphs and bootstrap equations in the ancillary files.

\subsection{Nonplanar contributions}\label{sec:nonplanar}

As stressed earlier, the cusp limit constraint and the graphical rule follow from general considerations of half-BPS correlators and apply to nonplanar contributions (or even general gauge groups). The bootstrap procedure is the same as in the planar limit, except that all $f$-graphs are now needed to construct the ansatz.

\begin{figure}[!ht]
    \centering
    \includegraphics[width=0.98\textwidth]{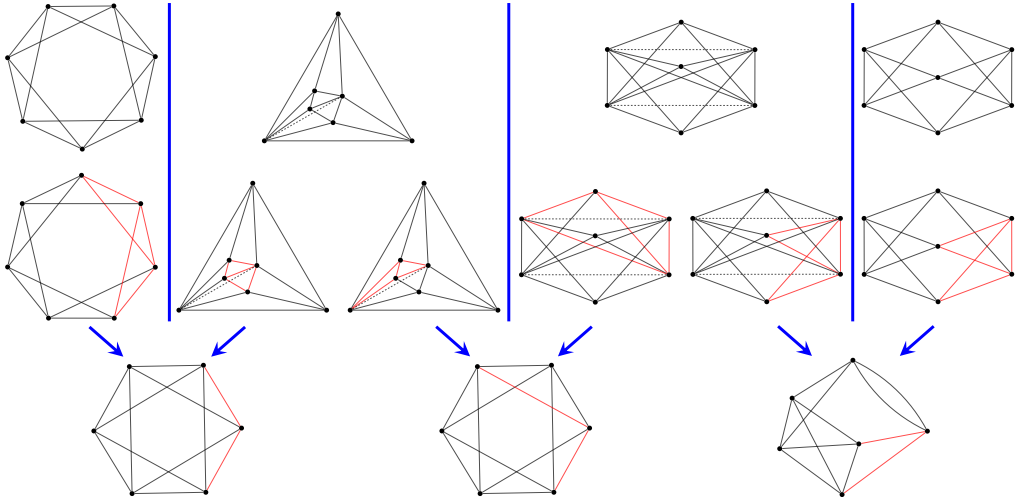}
    \caption{The graphical rule relating $F_7$ and $F_6$. The first row shows all 4 (non)planar $f$-graphs, which are ordered $f_{3}^{(7)},f_{1}^{(7)},f_{2}^{(7)},f_{4}^{(7)}$ for convenience of plotting. The second row shows all double-triangle-highlighted $f$-graphs $f^{\diamond(7)}$. Performing the $\mathcal P$ operation yields the cusp-highlighted $f^{\vee(6)}$ in the third row, where the first two arise from an allowed $f$-graph $f_{1}^{(6)}$ and the last one has a double pole. Carefully computing the symmetry factors yields the $7\to6$ bootstrap equations $\{c_3+c_1=b_1,\ c_1+c_2=b_1,\ c_2+\frac12c_4=0\}$.}
    \label{fig:nonplanar}
\end{figure}

Figure~\ref{fig:nonplanar} demonstrates the $7\to6$ bootstrap equations, where the graphical rule imposes three equations\footnote{Note that we always choose the $N_n$ planar $f$-graphs to be the first of the $\mathcal N_n$ (non-planar) $f$-graphs.} (after using the result from previous loop $b_1=1$):
\begin{equation}
    c_3+c_1=b_1=1,\quad c_1+c_2=b_1=1,\quad c_2+\frac12c_4=0.
\end{equation}
Solving the bootstrap equations yields
\begin{equation}
    c_1=1-\beta,\quad c_2=\beta,\quad c_3=\beta,\quad c_4=-2\beta.
\end{equation}
Up to the numbering of $f$-graphs, this is precisely eq.(4.7) of~\cite{Eden:2012tu}, where it was observed that the combination $-f_{1}^{(7)}+f_{2}^{(7)}+f_{3}^{(7)}-2f_{4}^{(7)}$ vanishes in four-dimensional spacetime due to Gram identities. We have also checked that solving the $8\to7$ bootstrap equations fixes all but $7=4(+3\text{ Gram})$ of the 32 coefficients, arriving at the same solution space as~\cite{Eden:2012tu}.

Solving the $9\to8$ bootstrap equations leaves a solution space of dimension 245 (out of 930), which may seem not particularly constraining. However, this is expected since the ansatz of $F_n$ including nonplanar contributions is the same as the ansatz of $F_n$ with general gauge groups. None of the conditions we used depend on the specific gauge group, and more constraints are needed that ``know about'' the gauge group (e.g., the twistor Feynman rules~\cite{Fleury:2019ydf}) in order to determine these nonplanar contributions.

\section{Discussions and Outlook}
\label{sec:discussions}
In this paper, we described a new graphical rule that constrains the loop integrand / Born-level correlator $F_n$ based on the cusp limit of half-BPS operators. The rule is so powerful that in the planar limit, it fixes $F_{\leq14}$ by itself and, together with the triangle rule, fixes $F_{15}$ as well, beating the previous record~\cite{Bourjaily:2016evz} by one loop/point. For the nonplanar contribution, the solution space of our graphical rule is found to coincide with~\cite{Eden:2012tu}. This immediately raises several questions, both technical and conceptual.

\paragraph{More efficient planar bootstrap}
We have seen in section~\ref{sec:catalan} that it is often worth studying a local subsystem of bootstrap equations, where we disproved the Catalan conjecture for $n=16$ although a complete $16\to15$ bootstrap is currently out of reach. It would be nice to find and study the local subsystem involving the invisible $f$-graphs such as chopped prisms, which would relate them to other visible $f$-graphs. For example, the 12-point chopped prism lives in the following simple local subsystem of $13\to12$ bootstrap equations:
\begin{gather*}
    \left.\begin{aligned}
    b\left(\includegraphics[scale=0.2,align=c]{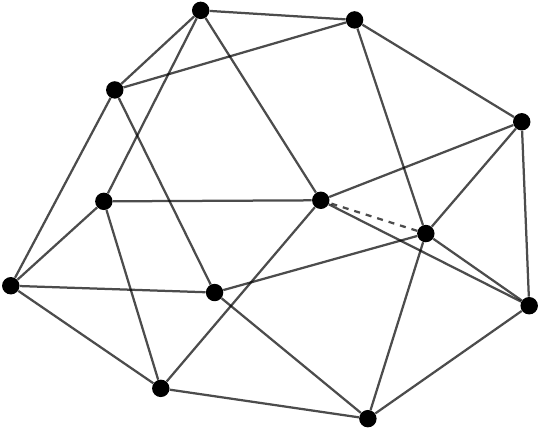}\right)&=c\left(\includegraphics[scale=0.2,align=c]{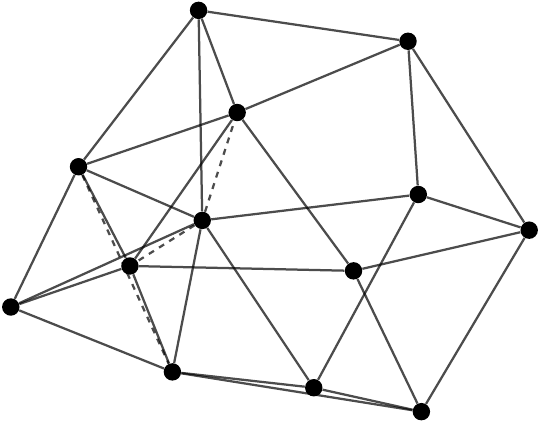}\right)\\
    b\left(\includegraphics[scale=0.2,align=c]{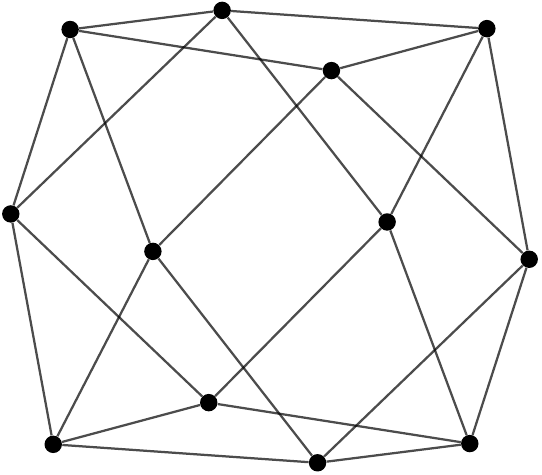}\right)&=2\ c\left(\includegraphics[scale=0.2,align=c]{newimages/local13_750994169761844669.pdf}\right)+2\ c\left(\includegraphics[scale=0.2,align=c]{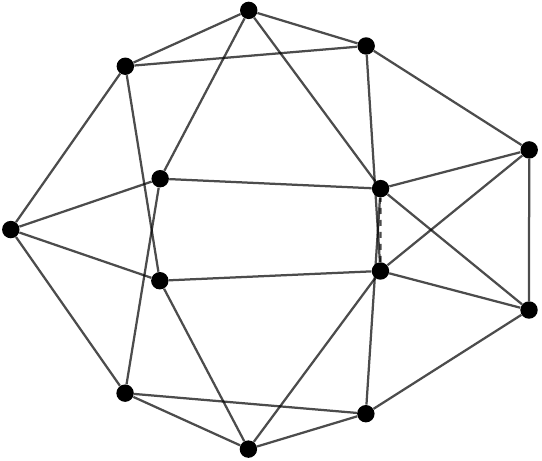}\right)\\
    b\left(\includegraphics[scale=0.2,align=c]{newimages/local12_6368767218444920853.pdf}\right)&=c\left(\includegraphics[scale=0.2,align=c]{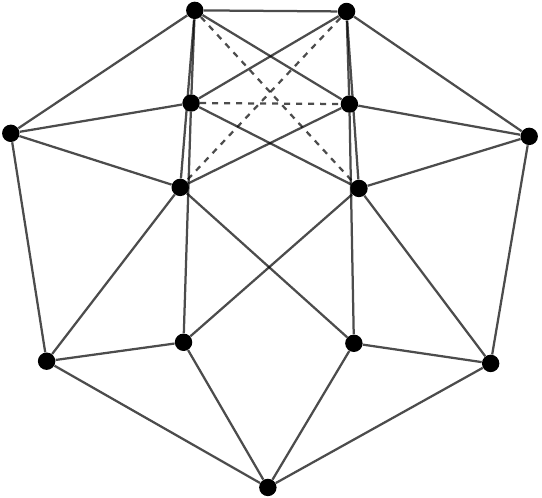}\right)+2\ c\left(\includegraphics[scale=0.2,align=c]{newimages/local13_5872184903479641311.pdf}\right)\\
    b\left(\includegraphics[scale=0.2,align=c]{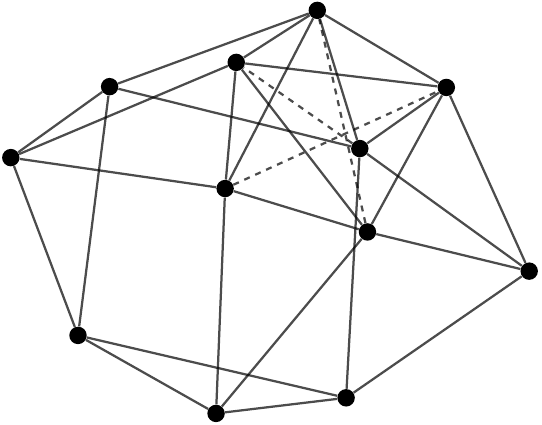}\right)&=c\left(\includegraphics[scale=0.2,align=c]{newimages/local13_5937507214248045616.pdf}\right)+c\left(\includegraphics[scale=0.2,align=c]{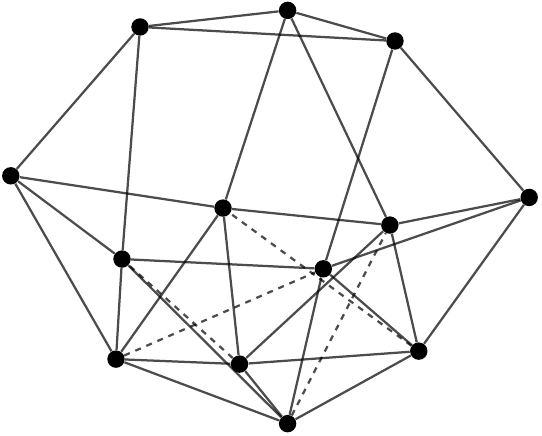}\right)\\
    b\left(\includegraphics[scale=0.2,align=c]{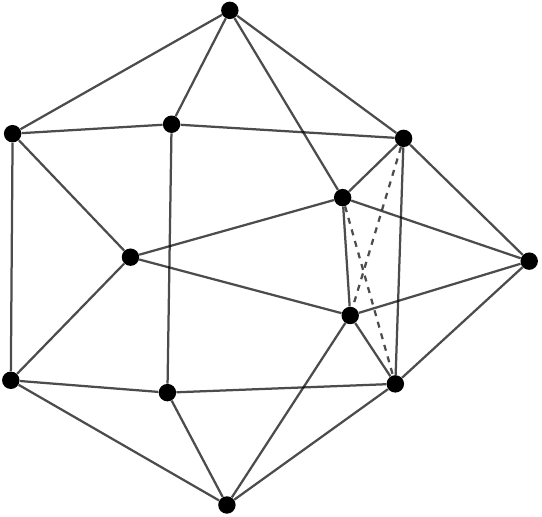}\right)&=c\left(\includegraphics[scale=0.2,align=c]{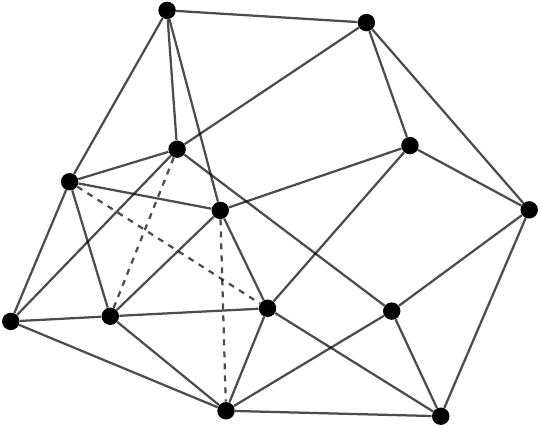}\right)+c\left(\includegraphics[scale=0.2,align=c]{newimages/local13_8767042035284913460.pdf}\right)\\
    b\left(\includegraphics[scale=0.2,align=c]{newimages/local12_4198257916268681920.pdf}\right)&=2\ c\left(\includegraphics[scale=0.2,align=c]{newimages/local13_653911436959188550.pdf}\right)
    \end{aligned}\right\}\\
    \implies b\left(\includegraphics[scale=0.2,align=c]{newimages/local12_6368767218444920853.pdf}\right)=-2\ b\left(\includegraphics[scale=0.2,align=c]{newimages/local12_4995902044211774480.pdf}\right)+2\ b\left(\includegraphics[scale=0.2,align=c]{newimages/local12_8418737201376832361.pdf}\right)-b\left(\includegraphics[scale=0.2,align=c]{newimages/local12_4198257916268681920.pdf}\right).
\end{gather*}
If one could find such a local subsystem for the 15-point chopped prism, it would settle the question ``Does our graphical rule suffice to fix the planar $F_{15}$, or do we really need other constraints such as the triangle rule?'' in less than two months. 

As noted in the main text, there are 619,981,403 $f$-graphs at $n=16$, posing significant technical challenges in terms of code efficiency, memory, and storage when advancing the bootstrap program to higher loops. One strategy to address this is to filter out as many non-contributing $f$-graphs as possible before applying the graphical rule. For example, at $n=14$, using the square rule (which yields one-to-one bootstrap equations) to find zero coefficients reduces the size of the ansatz by $81\%$, suggesting a potential efficiency improvement of about $70\%$ in the bootstrap process. Even more desirable is an efficient method to identify all zero $f$-graphs, which account for over $90\%$ of all planar graphs at $n=15$. We leave these questions for future investigation.

The fact that simple graphical rules uniquely determine the coefficients of all planar $f$-graphs strongly suggests that some hidden local properties of an $f$-graph may directly fix the value of its coefficient, independent of other $f$-graphs. Identifying these underlying properties is a complex task for humans, but well-suited for machine learning. It would be particularly interesting to explore the use of deep learning methods to determine these coefficients directly. There have already been several attempts to apply artificial intelligence to analytic calculations in theoretical high-energy physics~\cite{Alnuqaydan:2022ncd, Dersy:2022bym, Cai:2024znx, Cheung:2024svk}. Notably, the recent work by Cai et al.~\cite{Cai:2024znx} is analogous to our task, where the authors use a \emph{Transformer} model to predict the coefficients of symbol letters in the three-gluon form factor in $\mathcal{N}=4$ supersymmetric Yang-Mills theory. As a preliminary experiment, we conducted a binary classification task to train a simple encoder model (adapted from~\cite{kim2022pure}) to predict whether an $f$-graph coefficient is zero. After approximately 12 hours of training on a laptop with an \emph{Nvidia RTX 2060}, the model achieved an impressive $97.25\%$ accuracy at $n=13$, which encourages further exploration in this direction.

\paragraph{Correlator/amplitude duality and geometry}
It would also be interesting to study our graphical rule in the context of correlator/amplitude duality. Could it be the case that constraints from the square rule (and even pentagon rules {\it etc.}) always follow from our new graphical rule (in the planar limit)? Could we understand all these rules in the context of positive geometries such as the squared amplituhedron and the correlahedron~\cite{Dian:2021idl,Eden:2017fow,He:2024xed}? At least the soft limit of squared amplitudes (both tree and loop level), which are implied by our graphical rule, do follow from these geometric constructions.

\paragraph{All-loop characterization of $F_n$}
It is not inconceivable that the cusp limit and our graphical rule eventually suffice to determine the planar $F_n$ to all loops. After all, the $(n+1)\to n$ bootstrap equations completely fix $F_n$ up to $n=14$, and the $15\to14$ bootstrap equations almost fix $F_{15}$ except for a single term. It would be extremely interesting if an all-loop proof could be found using graph theory.

If an all-loop proof is too much to ask, more constraints are needed. For example, we have already seen that the triangle rule and our graphical rule nicely complement each other. Amusingly, the triangle rule follows from the OPE limit in Euclidean signature, while our graphical rule follows from the cusp limit in Minkowskian signature. Could there be another graphical rule that follows from certain limits in split signature? It would be very satisfying if such a graphical rule existed, and the relation and interplay between the three ``graphicateers'' would no doubt shed more light on the physics of half-BPS correlators in general. Another important constraint is the algebraic rule (e.g., eq.(5.19)) arising from the light-cone limit of~\cite{Eden:2012tu}. Naively, the cusp limit corresponds to the light-cone limit with the endpoint values $\alpha=0,1$, so we expect that the light-cone limit would be even more powerful than the cusp limit. But somehow the cusp limit turns out to be equally constraining for low loop orders $n\leq8$. Moreover, the cusp limit allows us to derive a graphical rule that is easily implemented, while the light-cone limit can only be described at the rational-function level at present, which quickly becomes impractical for higher loop orders. It would be interesting to see whether the light-cone limit allows for a nice graphical interpretation that utilizes the permutation invariance of $f$-graphs.

More generally, we have remarked that the nonplanar contributions cannot possibly be determined by these graphical rules that are agnostic to the specific gauge group. Therefore, it is natural to search for more constraints that know about the gauge group. Computationally, the twistor Feynman rules have been used in~\cite{Fleury:2019ydf} to successfully determine the nonplanar $F_8$. However, the computation was done on a graph-by-graph basis. Recent developments~\cite{Caron-Huot:2023wdh} suggest that half-BPS correlators in twistor space are closely related to geometric objects such as the $m=2$ amplituhedron~\cite{Arkani-Hamed:2013jha}. Such a geometric and global understanding could lead to new constraints on $F_n$ as a whole. Ultimately, the goal is to search for a complete characterization of $F_n$ to all loops, possibly applicable to general gauge groups.

\paragraph{More general observables}
The success of our graphical rule depends crucially on the permutation invariance of $F_n$, which is a special property of $\langle\mathcal O^4\mathcal L^{n-4}\rangle$. However, the general physical setup of the cusp limit has a much wider range of applicability. It would be very desirable to use these constraints to compute lower components $\langle\mathcal O^{\geq5}\mathcal L\cdots\mathcal L\rangle$ of the supercorrelator, or correlators involving higher Kaluza-Klein modes $\mathcal O_p\sim\mathop{\rm tr}(\phi^p)$. In particular, this would enable us to test and better understand the 10-dimensional symmetry observed in~\cite{Caron-Huot:2021usw,Caron-Huot:2023wdh}.

\begin{CJK*}{UTF8}{}
\CJKfamily{gbsn}
\acknowledgments
It is our pleasure to thank Jacob Bourjaily, Paul Heslop, Emeri Sokatchev, and Hua Xing Zhu (朱华星) for fruitful discussions. We also greatly appreciate Jacob Bourjaily for sharing his beautiful Mathematica codes which provided much inspiration. The work of SH is supported by the National Natural Science Foundation of China under Grant No. 12225510, 11935013, 12047503, 12247103, and by the New Cornerstone Science Foundation through the XPLORER PRIZE. The work of CS is supported by the China Postdoctoral Science Foundation under Grant No. 2022TQ0346, and the National Natural Science Foundation of China under Grant No. 12347146
\end{CJK*}

\appendix
\section{Details of the Lagrangian insertion procedure}\label{app:lagrangian}

In this appendix, we prove the crucial property of $\mathcal N=4$ SYM that the perturbative expansion of correlators can be computed by the (chiral, on-shell) Lagrangian insertion procedure. We strive to spell out all the details in one set of conventions, following the discussions in~\cite{Caron-Huot:2010ryg,Eden:2011yp} and references therein.

The $\mathcal N=4$ SYM action $S=\int{\rm d}^4x\,L(x)$ is defined by
\begin{equation}\label{eq:lagrangian}
    \begin{aligned}
        L&=\mathop{\rm tr}\left(-\frac12F_{\alpha\beta}F^{\alpha\beta}-\frac14\phi^{AB}[D_{\alpha\dot\alpha},[D^{\dot\alpha\alpha},\phi_{AB}]]+\frac{i}2\bar\psi_{A\dot\alpha}[D^{\dot\alpha\alpha},\psi^A_\alpha]-\frac{3i}2[D^{\dot\alpha\alpha},\bar\psi_{A\dot\alpha}]\psi^A_\alpha\right)\\
        &+\mathop{\rm tr}\left(\frac{g_{\rm YM}^2}8[\phi^{AB},\phi^{CD}][\phi_{AB},\phi_{CD}]-\sqrt2g_{\rm YM}\psi^{A\alpha}[\phi_{AB},\psi^B_\alpha]+\sqrt2g_{\rm YM}\bar\psi_{A\dot\alpha}[\phi^{AB},\bar\psi_B^{\dot\alpha}]\right).
    \end{aligned}
\end{equation}
Here, the gauge-covariant derivative $D_\mu:=\partial_\mu-ig_{\rm YM}A_\mu$ and the field strength $F_{\mu\nu}:=ig_{\rm YM}^{-1}[D_\mu,D_\nu]$. Our spinor and R-symmetry index conventions strictly follow section A.1 of~\cite{Eden:2011yp}. In particular,
\begin{equation}
    ig_{\rm YM}^{-1}[D_{\alpha\dot\alpha},D_{\beta\dot\beta}]=\epsilon_{\alpha\beta}\bar F_{\dot\alpha\dot\beta}+\epsilon_{\dot\alpha\dot\beta}F_{\alpha\beta}\quad\implies\quad F_{\alpha\beta}=-\frac i2g_{\rm YM}^{-1}\epsilon^{\dot\alpha\dot\beta}[D_{\alpha\dot\alpha},D_{\beta\dot\beta}].
\end{equation}
Under Hermitian conjugation, $(\psi^A_\alpha)^\dagger=\bar\psi_{A\dot\alpha}$, $(\phi^{AB})^\dagger=\phi_{AB}:=\frac12\varepsilon_{ABCD}\phi^{CD}$, and $(D_{\alpha\dot\beta})^\dagger=-D_{\beta\dot\alpha}$. Note that we have chosen to fix the total-derivative ambiguity of the Lagrangian in a specific way.

Denoting the matter fields collectively as $\Phi\equiv(\phi^{AB},\bar\psi_{A\dot\alpha},\psi^A_\alpha)$. A general gauge-invariant operator can be written as a functional $\mathcal O[D,\Phi]$ of the gauge connection and the matter fields. Consider the correlator  $\langle\mathcal O_{g_{\rm YM}}[D,\Phi]\rangle$ of gauge-invariant operators. We always normalize $\langle\mathcal O_{g_{\rm YM}}[D,\Phi]\rangle$ such that the perturbation series starts at $g_{\rm YM}^0$:
\begin{align}
    \langle\mathcal O_{g_{\rm YM}}[D,\Phi]\rangle&=\sum_{\ell=0}^\infty(g_{\rm YM}^2)^\ell\langle\mathcal O_{g_{\rm YM}}[D,\Phi]\rangle^{(\ell)},\\
    \langle\mathcal O_{g_{\rm YM}}[D,\Phi]\rangle^{(\ell)}&=\frac1{\ell!}\lim_{g_{\rm YM}\to0}\left(\frac{\rm d}{{\rm d}g_{\rm YM}^2}\right)^\ell\langle\mathcal O_{g_{\rm YM}}[D,\Phi]\rangle.\label{eq:pertseries}
\end{align}
In doing so, it is sometimes necessary for $\mathcal O_{g_{\rm YM}}[D,\Phi]$ to depend on the coupling explicitly, hence the subscript $_{g_{\rm YM}}$.

The correlator $\langle\mathcal O_{g_{\rm YM}}[D,\Phi]\rangle$ is defined by a Feynman path-integral:
\begin{equation}
    \langle\mathcal O_{g_{\rm YM}}[D,\Phi]\rangle=\int[\mathcal DD][\mathcal D\Phi]\,e^{iS[\Phi]}\mathcal O_{g_{\rm YM}}[D,\Phi].
\end{equation}
Introduce an auxiliary field $G_{\alpha\beta}$ to rewrite the path-integral:
\begin{gather}
    \langle\mathcal O_{g_{\rm YM}}[D,\Phi]\rangle=\int[\mathcal DD][\mathcal D\Phi][\mathcal DG]\,e^{i\hat S[\Phi,G]}\hat{\mathcal O}_{g_{\rm YM}}[D,\Phi,G],\quad\hat S=\int{\rm d}^4x\,\hat L(x),\\
    \hat L=\mathop{\rm tr}\left(-ig_{\rm YM}G_{\alpha\beta}F^{\alpha\beta}-\frac{g_{\rm YM}^2}2G_{\alpha\beta}G^{\alpha\beta}+(\text{same matter terms as }L)\right),
\end{gather}
where the new functional satisfies $\hat{\mathcal O}_{g_{\rm YM}}[D,\Phi,G_{\alpha\beta}=-ig_{\rm YM}^{-1}F_{\alpha\beta}]=\mathcal O_{g_{\rm YM}}[D,\Phi]$ on the support of the field equation of $G_{\alpha\beta}$. With the auxiliary field and the particular form of the kinetic terms in eq.\eqref{eq:lagrangian}\footnote{We write $\frac{\phi^{AB}}2$ and $\frac{G_{\alpha\beta}}2$ to avoid double-counting the degrees of freedom: $\frac{\phi^{AB}}2\frac{\delta}{\delta\phi^{AB}}\phi^{CD}=\phi^{CD}$ and $\frac{G_{\alpha\beta}}2\frac{\delta}{\delta G_{\alpha\beta}}G_{\gamma\delta}=G_{\gamma\delta}$. Later, when we integrate by parts, $\frac{\delta}{\delta\phi^{AB}}\frac{\phi^{AB}}2=6$ and $\frac{\delta}{\delta G_{\alpha\beta}}\frac{G_{\alpha\beta}}2=3$.},
\begin{equation}
    \hat L=\frac14\bar\psi_{A\dot\alpha}\frac{\delta\hat S}{\delta\bar\psi_{A\dot\alpha}}+\frac12\frac{\phi^{AB}}2\frac{\delta\hat S}{\delta\phi^{AB}}+\frac34\psi^A_\alpha\frac{\delta\hat S}{\delta\psi^A_\alpha}+\frac{G_{\alpha\beta}}2\frac{\delta\hat S}{\delta G_{\alpha\beta}}+\hat{\mathcal L},
\end{equation}
where the chiral, on-shell Lagrangian reads
\begin{equation}
    \hat{\mathcal L}=\mathop{\rm tr}\left(\frac{g_{\rm YM}^2}2G_{\alpha\beta}G^{\alpha\beta}+\sqrt2g_{\rm YM}\psi^{A\alpha}[\phi_{AB},\psi^B_\alpha]-\frac{g_{\rm YM}^2}8[\phi^{AB},\phi^{CD}][\phi_{AB},\phi_{CD}]\right). 
\end{equation}

Another important fact that we need is the homogeneity of fields. If we rescale the fields by defining $\phi=:g_{\rm YM}^{-1}\varphi$, $\psi=:g_{\rm YM}^{-1}\chi$, and $G=:g_{\rm YM}^{-2}g$, the gauge coupling factors out of the action. Collectively denoting $\hat\Phi\equiv(\bar\psi_{A\dot\alpha},\phi^{AB},\psi^A_\alpha,G_{\alpha\beta})$ and $\hat\Psi\equiv(\bar\chi_{A\dot\alpha},\varphi^{AB},\chi^A_\alpha,g_{\alpha\beta})$, we have $\hat S[\hat\Phi]=g_{\rm YM}^{-2}\hat s[\hat\Psi]$, where
\begin{equation}
    \begin{aligned}
        \hat s[\hat\Psi]&=\int{\rm d}^4x\mathop{\rm tr}\left(-\frac12\epsilon^{\dot\alpha\dot\beta}[D_{\alpha\dot\alpha},D_{\beta\dot\beta}]g^{\alpha\beta}-\frac12g_{\alpha\beta}g^{\alpha\beta}\right)\\
        &+\int{\rm d}^4x\mathop{\rm tr}\left(-\frac14\varphi^{AB}[D_{\alpha\dot\alpha},[D^{\dot\alpha\alpha},\varphi_{AB}]]+\frac{i}2\bar\chi_{A\dot\alpha}[D^{\dot\alpha\alpha},\chi^A_\alpha]-\frac{3i}2[D^{\dot\alpha\alpha},\bar\chi_{A\dot\alpha}]\chi^A_\alpha\right)\\
        &+\int{\rm d}^4x\mathop{\rm tr}\left(\frac18[\varphi^{AB},\varphi^{CD}][\varphi_{AB},\varphi_{CD}]-\sqrt2\chi^{A\alpha}[\varphi_{AB},\chi^B_\alpha]+\sqrt2\bar\chi_{A\dot\alpha}[\varphi^{AB},\bar\chi_B^{\dot\alpha}]\right).
    \end{aligned}
\end{equation}
The measure of the path-integral transforms under the rescaling\footnote{Recall that bosonic and fermionic differentials transform inversely.}:
\begin{equation}
    [\mathcal D\phi]=g_{\rm YM}^{-6\mathcal N}[\mathcal D\varphi],\quad[\mathcal D\bar\psi][\mathcal D\psi]=g_{\rm YM}^{16\mathcal N}[\mathcal D\bar\chi][\mathcal D\chi],\quad[\mathcal DG]=g_{\rm YM}^{-2\times3\mathcal N}[\mathcal Dg],
\end{equation}
where $\mathcal N$ is an infinite constant, which can be interpreted as the dimension of the gauge group (e.g., $SU(N_c)$) times the volume of spacetime. We will be concerned with correlators which are homogeneous with respect to the rescaled fields:
\begin{equation}
    \hat{\mathcal O}_{g_{\rm YM}}[D,\hat\Phi]=g_{\rm YM}^{-m}\hat{\mathcal O}[D,\hat\Psi].
\end{equation}
Note that $\hat{\mathcal O}[D,\hat\Psi]$ has no explicit dependence on $g_{\rm YM}$ and therefore no subscript. For example\footnote{Recall that the overall $g_{\rm YM}^{-2q}$ is needed for the correlator $\langle\hat{\mathcal O}^{(p,q)}_{g_{\rm YM}}[D,\hat\Phi]\rangle$ to start at $g_{\rm YM}^0$.},
\begin{gather}
    \hat{\mathcal O}^{(p,q)}_{g_{\rm YM}}[D,\hat\Phi]:=g_{\rm YM}^{-2q}\left[\mathop{\rm tr}(\phi^2)\right]^p\hat{\mathcal L}^q=g_{\rm YM}^{-2p-4q}\left[\mathop{\rm tr}(\varphi^2)\right]^p\hat{\boldsymbol\ell}^q,\label{eq:mValue}\\
    \hat{\mathcal L}(\hat\Phi)=g_{\rm YM}^{-2}\hat{\boldsymbol\ell}(\hat\Psi)=g_{\rm YM}^{-2}\mathop{\rm tr}\left(\frac12g_{\alpha\beta}g^{\alpha\beta}+\sqrt2\chi^{A\alpha}[\varphi_{AB},\chi^B_\alpha]-\frac18[\varphi^{AB},\varphi^{CD}][\varphi_{AB},\varphi_{CD}]\right).
\end{gather}

We are now ready to compute the perturbative expansion of $\langle\mathcal O_{g_{\rm YM}}[D,\Phi]\rangle$. Consider the following derivative written in terms of the rescaled fields,
\begin{align}
    &g_{\rm YM}^2\frac{\rm d}{{\rm d}g_{\rm YM}^2}\langle\mathcal  O_{g_{\rm YM}}[D,\Phi]\rangle\nonumber\\
    =\,&g_{\rm YM}^2\frac{\rm d}{{\rm d}g_{\rm YM}^2}\int[\mathcal DD]g_{\rm YM}^{-6\mathcal N}[\mathcal D\varphi]g_{\rm YM}^{16\mathcal N}[\mathcal D\bar\chi][\mathcal D\chi]g_{\rm YM}^{-2\times3\mathcal N}[\mathcal Dg]\,e^{ig_{\rm YM}^{-2}\hat s[\hat\Psi]}g_{\rm YM}^{-m}\hat{\mathcal O}[D,\hat\Psi]\nonumber\\
    =\,&-\frac{6-16+2\times3}2\mathcal N\langle\hat{\mathcal O}_{g_{\rm YM}}[D,\hat\Phi]\rangle-\frac m2\langle\hat{\mathcal O}_{g_{\rm YM}}[D,\hat\Phi]\rangle-i\langle\hat S[\hat\Phi]\hat{\mathcal O}_{g_{\rm YM}}[D,\hat\Phi]\rangle\nonumber\\
    =\,&-\frac{6-16+2\times3}2\mathcal N\langle\hat{\mathcal O}_{g_{\rm YM}}[D,\hat\Phi]\rangle-\frac m2\langle\hat{\mathcal O}_{g_{\rm YM}}[D,\hat\Phi]\rangle-i\int{\rm d}^4x\langle\hat L(x)\hat{\mathcal O}_{g_{\rm YM}}[D,\hat\Phi]\rangle.\label{eq:qdq}
\end{align}
Consider the part in $\hat L$ proportional to the field equations. For each field component $\hat\Phi$,
\begin{align}
    &\int{\rm d}^4x\left\langle\hat\Phi(x)\frac{\delta\hat S[\hat\Phi]}{\delta\hat\Phi(x)}\hat{\mathcal O}_{g_{\rm YM}}[D,\hat\Phi]\right\rangle\nonumber\\
    =\,&\int{\rm d}^4x\int[\mathcal DD][\mathcal D\hat\Phi]\,e^{i\hat S[\hat\Phi]}\hat\Phi(x)\frac{\delta\hat S[\hat\Phi]}{\delta\hat\Phi(x)}\hat{\mathcal O}_{g_{\rm YM}}[D,\hat\Phi]\nonumber\\
    =\,&-i\int{\rm d}^4x\int[\mathcal DD][\mathcal D\hat\Phi]\,\hat\Phi(x)\frac{\delta e^{i\hat S[\hat\Phi]}}{\delta\hat\Phi(x)}\hat{\mathcal O}_{g_{\rm YM}}[D,\hat\Phi]\nonumber\\
    =\,&i\int{\rm d}^4x\int[\mathcal DD][\mathcal D\hat\Phi]\,e^{i\hat S[\hat\Phi]}\left((-)^{{\rm sgn}(\hat\Phi)}\frac{\delta\hat\Phi(x)}{\delta\hat\Phi(x)}\hat{\mathcal O}_{g_{\rm YM}}[D,\hat\Phi]+\hat\Phi(x)\frac{\delta\hat{\mathcal O}_{g_{\rm YM}}[D,\hat\Phi]}{\delta\hat\Phi(x)}\right).\nonumber\\
    =\,&i(-)^{{\rm sgn}(\hat\Phi)}\mathcal N\langle\hat{\mathcal O}_{g_{\rm YM}}[D,\hat\Phi]\rangle+i\left\langle\int{\rm d}^4x\,\hat\Phi(x)\frac{\delta\hat{\mathcal O}_{g_{\rm YM}}[D,\hat\Phi]}{\delta\hat\Phi(x)}\right\rangle.\nonumber
\end{align}
Here, ${\rm sgn}(\hat\Phi)=0$ or $1$ depending on whether $\hat\Phi$ is bosonic or fermionic. Counting components,
\begin{align}
    &\int{\rm d}^4x\langle\hat L(x)\hat{\mathcal O}_{g_{\rm YM}}[D,\hat\Phi]\rangle\nonumber\\
    =\,&i\left(\frac14\times(-8)+\frac12\times6+\frac34\times(-8)+1\times3\right)\mathcal N\langle\hat{\mathcal O}_{g_{\rm YM}}[D,\hat\Phi]\rangle\label{eq:cancelN}\\
    +\,&i\left\langle\int{\rm d }^4 x \left(\frac14\bar\psi_{A\dot\alpha}\frac{\delta}{\delta\bar\psi_{A\dot\alpha}}+\frac12\frac{\phi^{AB}}2\frac{\delta}{\delta\phi^{AB}}+\frac34\psi^A_\alpha\frac{\delta}{\delta\psi^A_\alpha}+\frac{G_{\alpha\beta}}2\frac{\delta}{\delta G_{\alpha\beta}}\right)\hat{\mathcal O}_{g_{\rm YM}}[D,\hat\Phi]\right\rangle\label{eq:cancelM}\\
    +\,&\int{\rm d}^4x\langle\hat{\mathcal L}(x)\hat{\mathcal O}_{g_{\rm YM}}[D,\hat\Phi]\rangle.
\end{align}
Plugging into eq.\eqref{eq:qdq}, we see that the field equation terms eq.\eqref{eq:cancelN} cancel the first term in eq.\eqref{eq:qdq}. At least for the special class of correlators $\langle\hat{\mathcal O}^{(p,q)}_{g_{\rm YM}}[D,\hat\Phi]\rangle$, evaluating eq.\eqref{eq:cancelM} yields $i(p+2q)\langle\hat{\mathcal O}^{(p,q)}_{g_{\rm YM}}[D,\hat\Phi]\rangle$, which cancels the second term in eq.\eqref{eq:qdq} because $m=2p+4q$ in this case (eq.\eqref{eq:mValue}). In the end,
\begin{equation}
    \frac{\rm d}{{\rm d}g_{\rm YM}^2}\langle\hat{\mathcal O}^{(p,q)}_{g_{\rm YM}}[D,\hat\Phi]\rangle=-i\int{\rm d}^4x_{p+q+1}\left\langle\hat{\mathcal O}^{(p,q)}_{g_{\rm YM}}[D,\hat\Phi]\frac{\hat{\mathcal L}(x_{p+q+1})}{g_{\rm YM}^2}\right\rangle.
\end{equation}
Using eq.\eqref{eq:pertseries} and integrating out the auxiliary field $G_{\alpha\beta}$, we obtain the desired result:
\begin{equation}\label{eq:insertion}
    \langle\mathcal O^{(p,q)}_{g_{\rm YM}}[D,\Phi]\rangle^{(\ell)}=\frac{(-i)^\ell}{\ell!}\int{\rm d}^4x_{p+q+1}\cdots{\rm d}^4x_{p+q+\ell}\left\langle\mathcal O^{(p,q)}_{g_{\rm YM}}[D,\Phi]\frac{\mathcal L(x_{p+q+1})}{g_{\rm YM}^2}\cdots\frac{\mathcal L(x_{p+q+\ell})}{g_{\rm YM}^2}\right\rangle^{(0)},
\end{equation}
where
\begin{equation}
    \mathcal L=\mathop{\rm tr}\left(-\frac12F_{\alpha\beta}F^{\alpha\beta}+\sqrt2q\psi^{A\alpha}[\phi_{AB},\psi^B_\alpha]-\frac{q^2}8[\phi^{AB},\phi^{CD}][\phi_{AB},\phi_{CD}]\right).
\end{equation}
Since $\langle\mathcal L\cdots\mathcal L\rangle=0$ due to superconformal symmetry~\cite{Eden:2011yp}, every inserted Lagrangian must be connected to $\mathcal O_{g_{\rm YM}}^{(p,q)}[D,\Phi]$, establishing eq.\eqref{eq:insertion} for the connected correlators:
\begin{equation}
    \langle\mathcal O^{(p,q)}_{g_{\rm YM}}[D,\Phi]\rangle_{\rm conn}^{(\ell)}=\frac{(-i)^\ell}{\ell!}\int{\rm d}^4x_{p+q+1}\cdots{\rm d}^4x_{p+q+\ell}\left\langle\mathcal O^{(p,q)}_{g_{\rm YM}}[D,\Phi]\frac{\mathcal L(x_{p+q+1})}{g_{\rm YM}^2}\cdots\frac{\mathcal L(x_{p+q+\ell})}{g_{\rm YM}^2}\right\rangle_{\rm conn}^{(0)},
\end{equation}
\section{Essentially one-loop argument of the triangle rule}\label{app:triangle}

In this appendix, we provide an alternative argument of the triangle rule to complement the one given in~\cite{Bourjaily:2016evz}. In parallel with our rule described in the main text, our argument of the triangle rule is essentially one-loop, thus avoiding the analysis of complicated sub-divergences in multi-loop integrals, which are needed for an argument based on the logarithm of the correlator.

We start with the Lagrangian insertion relation eq.~\eqref{eq:insertionM} in Minkowski spacetime:
\begin{align}
    \langle\mathcal O_1\bar{\mathcal O}_2\mathcal O_3\bar{\mathcal O}_4&(g_{\rm YM}^{-2}\mathcal L_5)\cdots(g_{\rm YM}^{-2}\mathcal L_n)\rangle_{\rm conn}^{(\ell)}\nonumber\\
    &=\frac{(-i)^\ell}{\ell!}\int{\rm d}^4x_{n+1}\cdots{\rm d}^4x_{n+\ell}\langle\mathcal O_1\bar{\mathcal O}_2\mathcal O_3\bar{\mathcal O}_4(g_{\rm YM}^{-2}\mathcal L_5)\cdots(g_{\rm YM}^{-2}\mathcal L_{n+\ell})\rangle_{\rm conn}^{(0)},\\
    \langle\mathcal O_1\bar{\mathcal O}_2\mathcal O_3\bar{\mathcal O}_4&(a^{-1}\mathcal L_5)\cdots(a^{-1}\mathcal L_{4+\ell})\rangle_{\rm conn}^{(0)}=\frac{N_c^2-1}{(4\pi^2)^{4+\ell}}\frac1{x_{12}^2x_{23}^2x_{34}^2x_{14}^2}2\xi_4F_{4+\ell}(x_1,\cdots,x_{4+\ell}).
\end{align}
Since the triangle rule follows from the OPE limit in Euclidean spacetime, hereafter in this appendix, we will work with Euclidean signature. In order to simplify notation, we will not use a different notation $x$ for points in Euclidean spacetime. Using ${\rm d}^4x^{(M)}=-i{\rm d}^4x^{(E)}$:
\begin{align}
    \langle\mathcal O_1\bar{\mathcal O}_2\mathcal O_3\bar{\mathcal O}_4&(g_{\rm YM}^{-2}\mathcal L_5)\cdots(g_{\rm YM}^{-2}\mathcal L_n)\rangle_{\rm conn}^{(\ell)}\nonumber\\
    &=\frac{(-)^\ell}{\ell!}\int{\rm d}^4x_{n+1}\cdots{\rm d}^4x_{n+\ell}\langle\mathcal O_1\bar{\mathcal O}_2\mathcal O_3\bar{\mathcal O}_4(g_{\rm YM}^{-2}\mathcal L_5)\cdots(g_{\rm YM}^{-2}\mathcal L_{n+\ell})\rangle_{\rm conn}^{(0)},\\
    \langle\mathcal O_1\bar{\mathcal O}_2\mathcal O_3\bar{\mathcal O}_4&(a^{-1}\mathcal L_5)\cdots(a^{-1}\mathcal L_{4+\ell})\rangle_{\rm conn}^{(0)}=\frac{N_c^2-1}{(4\pi^2)^{4+\ell}}\frac1{x_{12}^2x_{23}^2x_{34}^2x_{14}^2}2\xi_4F_{4+\ell}(x_1,\cdots,x_{4+\ell}).
\end{align}
Let us consider the OPE limit $x_2\to x_1$:
\begin{equation}
    \mathcal O(x_1)\bar{\mathcal O}(x_2)=\text{contribution of }\mathbf1\text{ and }O(x_1,t)+\frac{c_\mathcal K(a)}{(x_{12}^2)^{1-\frac12\gamma_\mathcal K(a)}}\mathcal K(x_1)+\text{regular}.
\end{equation}
Here, $\mathcal K=\mathop{\rm tr}(\phi^I\phi^I)$ is the Konishi operator. Applying the OPE for the correlator of interest,
\begin{equation}\label{eq:tri-master}
    \begin{aligned}
        &\langle\mathcal O_1\bar{\mathcal O}_2\mathcal O_3\bar{\mathcal O}_4(a^{-1}\mathcal L_5)\cdots(a^{-1}\mathcal L_n)\rangle_{\rm conn}\\
        =\,&\frac{c_\mathcal K(a)}{(x_{12}^2)^{1-\frac12\gamma_\mathcal K(a)}}\langle\mathcal K_1\mathcal O_3\bar{\mathcal O}_4(a^{-1}\mathcal L_5)\cdots(a^{-1}\mathcal L_n)\rangle_{\rm conn}+\text{regular}.
    \end{aligned}
\end{equation}
Note that the identity and 1/2-BPS contributions vanish. This is because the Grassmann degrees of $\langle OO\mathcal L\cdots\mathcal L\rangle$ and $\langle OOO\mathcal L\cdots\mathcal L\rangle$ in the stress tensor supercorrelator are too high.

Consider the ratio of the order-$a^1$ and the order-$a^0$ contributions of the above equation. On the right-hand side, the Konishi anomalous dimension leads to a single-log divergence:
\begin{equation}
    \frac{\text{RHS of eq.\eqref{eq:tri-master}}|_{a^1}}{\text{RHS of eq.\eqref{eq:tri-master}}|_{a^0}}=\frac{\gamma_\mathcal K|_{a^1}}2\log x_{12}^2+\text{other}.
\end{equation}
In particular, the ``other'' terms do not contribute to the $\log x_{12}^2$ divergence. On the left-hand side, since loop corrections are computed with Lagrangian insertion,
\begin{equation}
    \frac{\text{LHS of eq.\eqref{eq:tri-master}}|_{a^1}}{\text{LHS of eq.\eqref{eq:tri-master}}|_{a^0}}=\frac1{-4\pi^2}\frac{\int{\rm d}^4x_{n+1}^{(E)}\,F_{n+1}}{F_n}.
\end{equation}
Putting the pieces together and moving around some terms, we get
\begin{equation}\label{eq:tri-div}
    \left.\frac1{-4\pi^2}\lim_{x_2\to x_1}\int{\rm d}^4x_{n+1}^{(E)}\,x_{12}^2F_{n+1}\right|_{\substack{\text{single-log}\\\text{divergence}}}=\frac{\gamma_\mathcal K|_{a^1}}2\log x_{12}^2\times\lim_{x_2\to x_1}x_{12}^2F_n.
\end{equation}
On the left-hand side, we have a one-loop Feynman integral and the single-log divergence can only arise from the region $x_{n+1}\approx x_1$. We can extract the divergence as follows:
\begin{align*}
    &\text{LHS of eq.\eqref{eq:tri-div}}\\
    =\,&\frac1{-4\pi^2}\left.\lim_{x_2\to x_1}\int\frac{{\rm d}^4x_{n+1}^{(E)}}{x_{1,n+1}^2x_{2,n+1}^2}\right|_{\substack{\text{single-log}\\\text{divergence}}}\times\lim_{x_2,x_{n+1}\to x_1}\,x_{12}^2x_{1,n+1}^2x_{2,n+1}^2F_{n+1}\\
    =\,&\frac14\log x_{12}^2\times\lim_{x_2,x_{n+1}\to x_1}\,x_{12}^2x_{1,n+1}^2x_{2,n+1}^2F_{n+1}.
\end{align*}
Therefore, we obtain the OPE limit constraint on $F_n$:
\begin{equation}
    \lim_{x_2,x_{n+1}\to x_1}\,x_{12}^2x_{1,n+1}^2x_{2,n+1}^2F_{n+1}=6\lim_{x_2\to x_1}x_{12}^2F_n,
\end{equation}
where we have used $\gamma_\mathcal K|_{a^1}=3$, which could be obtained either from eq.(1.1) of~\cite{Eden:2012fe}, or by plugging in the $n=5$ result eq.\eqref{eq:n56result}. This is precisely eq.(3.15) of~\cite{Bourjaily:2016evz}, from which the triangle rule can be derived.

\bibliographystyle{JHEP}
\bibliography{inspire.bib}

\end{document}